\documentclass{egpubl}
\usepackage{eurovis2021}

\SpecialIssuePaper         

\usepackage[T1]{fontenc}
\usepackage{dfadobe}

\usepackage{cite}  
\BibtexOrBiblatex
\electronicVersion
\PrintedOrElectronic
\ifpdf \usepackage[pdftex]{graphicx} \pdfcompresslevel=9
\else \usepackage[dvips]{graphicx} \fi

\usepackage[most]{tcolorbox}

\usepackage{egweblnk}
\usepackage{newtxmath,tikz-cd}

\title[What are Table Cartograms Good for Anyway? An Algebraic Analysis]%
      {What are Table Cartograms Good for Anyway? \\An Algebraic Analysis}

\author[A. McNutt]
{\parbox{\textwidth}{\centering A. McNutt \orcid{0000-0001-8255-4258}
        }
        \\
{\parbox{\textwidth}{\centering Department of Computer Science, University of Chicago
       }
}
}

\begin{document}

\maketitle
\newcommand{\osf}{https://osf.io/5r6fn/?view_only=c8708b2cee93416284df7c6a45892aee}
\newcommand{\etal}{et al.\@ }
\newcommand{\etals}{et al.\@'s }
\newcommand{\midsepremove}{\aboverulesep = 0.0mm \belowrulesep = 0.0mm}
\newcommand{\midsepdefault}{\aboverulesep = 0.605mm \belowrulesep = 0.984mm}

\newcommand\am[1]{{\color{red}[AM: #1]}}

\newcommand{\parahead}[1]{\altparahead{#1.}}

\newcommand{\altparahead}[1]
{%
    \vspace{0.07in}%
    \noindent%
    \textbf{\textit{#1}}%
}
\newcommand{\secref}[1]{\hyperref[#1]{Sec.~\ref*{#1}}}
\newcommand{\noref}[1]{\hyperref[#1]{~\ref*{#1}}}
\newcommand{\appendixref}[1]{\hyperref[#1]{Appendix.~\ref*{#1}}}
\newcommand{\figref}[1]{\hyperref[#1]{Fig.~\ref*{#1}}}
\newcommand{\eqnref}[1]{\hyperref[#1]{Eqn.~\ref*{#1}}}
\newcommand{\tabref}[1]{\hyperref[#1]{Table ~\ref*{#1}}}

\newcommand{\task}[1]{\textsc{#1}}

\newcommand{\taco}{TACO}

\newcommand{\confuser}{\textbf{\emph{Confuser}}}
\newcommand{\confusers}{\textbf{\emph{Confusers}}}
\newcommand{\halluc}{\textbf{\emph{Hallucinator}}}
\newcommand{\hallucs}{\textbf{\emph{Hallucinators}}}

\newcommand{\spaceit}{\vspace{-0.2in}}

\begin{abstract}
    Unfamiliar or esoteric visual forms arise in many areas of visualization.
    While such forms can be intriguing, it can be unclear how to make effective use of them without long periods of practice or costly user studies.
    In this work we analyze the table cartogram---a graphic which visualizes tabular data by bringing the areas of a grid of quadrilaterals into correspondence with the input data, like a heat map that has been \emph{``area-ed''} rather than colored.
    Despite having existed for several years, little is known about its appropriate usage.
    We mend this gap by using Algebraic Visualization Design to show that they are best suited to relatively small tables with ordinal axes for some comparison and outlier identification tasks.
    In doing so we demonstrate a discount theory-based analysis that can be used to cheaply determine best practices for unknown visualizations.

    \ccsdesc[500]{Human-centered computing~Visualization design and evaluation methods}
    \ccsdesc[500]{Human-centered computing~Visualization theory, concepts and paradigms}

    \printccsdesc
\end{abstract}

\section{Introduction}

\maketitle

Understanding whether a chart has been used effectively is an important problem in the practice of data visualization.
Possessing a clear notion of what is \emph{``good''} or \emph{``bad''} usage is critical, as it can guide designers towards impactful information-rich graphics and away from deceptive displays \cite{blackhat}.
Yet even for those familiar with visualization best practices, it can be difficult to know if an unfamiliar chart type has been used effectively.

There are at least a dozen decades of advice \cite{wickham2013graphical}
on how to best use extant chart forms for known data types, yet this advice rarely applies to novel charts.
This lack of guidance can impede effective usage by practitioners, can cause domain experts to question a design's validity \cite{van2018philosophical}, and can impede automated analyses from making relevant suggestions\cite{mcnutt2020divining}.
Developing a technical understanding of effective usage for a novel chart can be a quagmire of disentangling aesthetic and novelty responses\cite{cawthon2007effect}  through often slow or  costly user studies\cite{adar2020communicative}.

A potential salve is to utilize a theory-based analysis to generate guidelines.
This would enable users without expertise or access to specialized analysis software to generate best practices for themselves.
This is similar in spirit to discount usability studies\cite{zuk2006heuristics}, which succinctly characterize a system's usability through evaluation of heuristics by a small number of analysts.
Prior efforts to apply theory in this way have focused on guiding design processes already situated in their task and context, rather than on understanding usage of a particular chart form \cite{kindlmann2014algebraic, zuk2006theoretical}.

We demonstrate the potential of such an analysis by investigating the properties of an uncommon visualization---the table cartogram (\taco{})---
through a lens informed by
Kindlmann and Scheidegger's
Algebraic Visualization Design (AVD)\cite{kindlmann2014algebraic}. We study \taco{}s because, despite having existed for several years\cite{evans2013table}, appropriate usage is still unknown \cite{mcnutt2020Minimally}.
We focus on AVD because---in contrast to other frameworks---it provides concrete assertions about visualization quality which are human-operable and interpretable.
Further, its methods are removed from the embodied response to visualizations, reducing the novelty effect that might be brought on by unusual visual forms.

We contribute an example of how AVD might be used to derive guidelines for novel visual forms, by considering what data (\secref{sec:data-analysis}) and tasks (\secref{sec:task-analysis}) are appropriate for one such form.
While not every question can be answered using these tools
(those related to perceptual quality are typically out of reach),
we are able to construct cogent guidelines for most basic usage questions.
This allows us to contribute recommendations on how \taco{}s might be effectively used.
We argue that they are well suited to some comparison and outlier identification tasks for flat tables that have ordinal rows and columns.
We suggest that they may be effective in contexts in which analytical insight is not the primary goal, as well as in discrete representations of time---such as month calendars.
While not every property will be surprising to those familiar with similar graphics, our investigation offers a full picture of \taco{}s that explains their usage and helps form an agenda for their future study.

\section{Related Work}

Our work constructs a theory-driven analysis of an uncommon graphic, the table cartogram. We now ground this analysis by describing \taco{}'s history and known properties, then prior theory-based evaluations, and finally our theory of interest, AVD.

\subsection{Table Cartograms}\label{sec:table-cartograms}

\emph{Table cartograms} (\taco{}s) are a specialized variation of value-by-area maps (cartograms)
that shows tabular data rather than solely geographic data~\cite{evans2013table}.
They depict a table of positive numbers as a grid of quadrilaterals, constrained to a rectangle, whose areas are brought into correspondence with the input data.
The visual effect is that of a shaded matrix that has been \emph{``area-ed''} rather than colored.
They can be characterized as having \emph{Planar Grid Topology} (none of the cells overlap) and an \emph{Accurate Data Embedding} (data is represented accurately as area) \cite{mcnutt2020Minimally}.
While color is a consonant secondary encoding, its use is not definitionally required.
Inoue and Li\cite{inoue2020optimization} refer to \taco{}s not constrained to a containing rectangle as \emph{deformed table cartograms}.
Although these are of interest, we instead focus on \emph{undeformed table cartograms} as constraining the chart unambiguously defines the meaning of area and usefully limits the domain of our analysis.

\begin{figure}[ht] 
      \centering
      \includegraphics[width=\columnwidth]{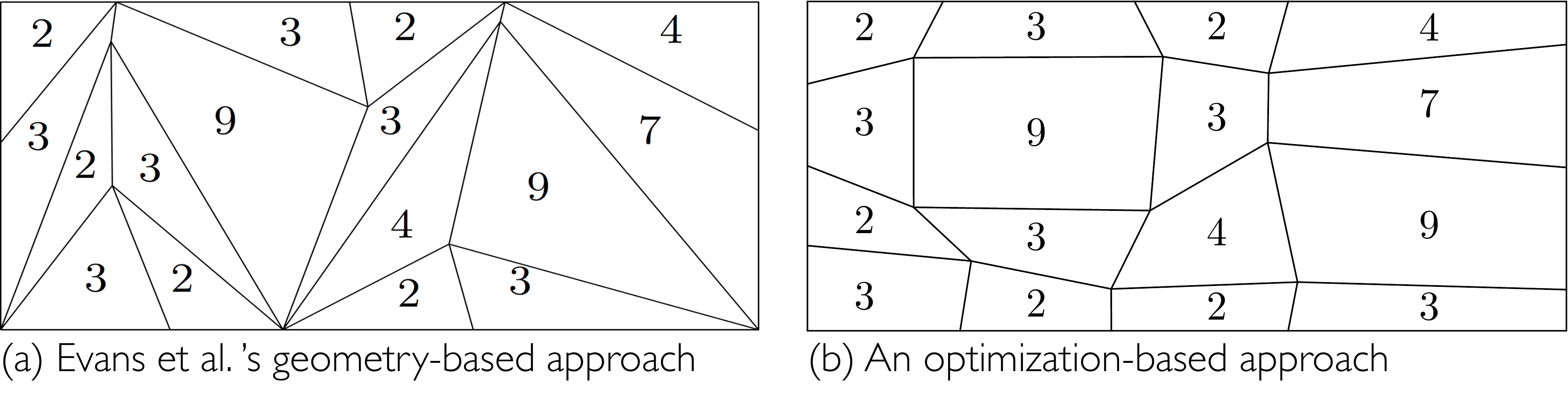}
      \caption{
            Table cartograms admit multiple equally-accurate layouts for a given input.
            This is a \halluc{}:
            the multiplicity of correct solutions may yield varying interpretations.
      }
      \label{fig:apples-to-apples}
      \spaceit
\end{figure}

\taco{}s were first described by Evans \etal \cite{evans2013table}. They use a computational geometry-based algorithm to constructively demonstrate that all tables of positive numbers admit a \taco{}.
Subsequent work\cite{li2019table, inoue2020optimization, mcnutt2020Minimally} found that more expressive graphics could be produced with optimization-based techniques, utilizing the fact the problem is under-constrained\cite{inoue2020optimization}. This allows for multiple outputs for a single input, as in \figref{fig:apples-to-apples}.

Prior studies primarily focus on \taco{}'s construction without providing substantial considerations on how they might be usefully employed.
Evans \etal \cite{evans2013table} explore a series of designs, but provide no usage guidance.
Inoue and Li\cite{inoue2020optimization} briefly touch on usage (which informs our discussion of data types in \secref{sec:data-analysis}), however they do not consider task effectiveness or appropriate data domains.
Confounding the development of usage guidelines is the common identification of \taco{}s as a form of geographic cartogram\cite{alam2015quantitative, nusrat2016state}.
While they can be used for such data, few geographies have topologies that can be mapped to a grid without substantial adjacency distortion (grid mapping schemes\cite{eppstein2015improved}
can reduce these distortions\cite{nusrat2016state}).
We focus instead on tabular data as it covers tabularized geographic data and is rarely supported among familiar chart forms (tables and shaded-matrices being the best-known exceptions).
Despite these prior studies, little is known about effective \taco{} usage.

In this paper we seek to rectify this gap in understanding by answering the question: \emph{What are table cartograms good for anyway?}

\definecolor{hallucColor}{HTML}{F7798B}
\definecolor{symColor}{HTML}{B2EC84}

\tcbset{on line,
    boxsep=4pt, left=0pt,right=0pt,top=0pt,bottom=0pt,
    colframe=white,colback=hallucColor,
    highlight math style={enhanced}
}

\begin{figure}[t]
    \centering
    \includegraphics[width=\columnwidth]{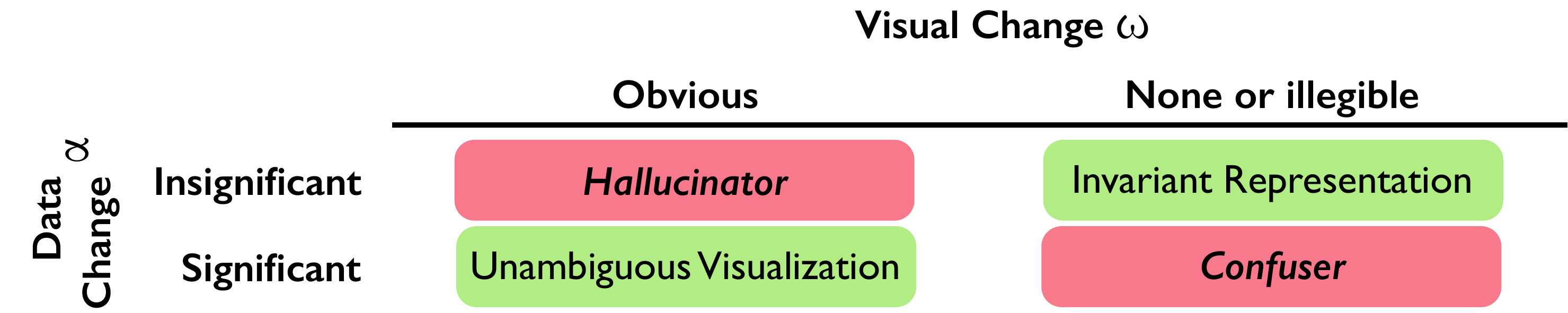}
    \vspace{-0.2in}
    \caption{
        \label{fig:avd-error-modes}
        The primary \tcbox[colback=symColor]{success} and \tcbox{failure} modes in AVD.
    }
    \vspace{-0.3in}
\end{figure}

\subsection{Visualization Analysis}\label{sec:theory}

A number of theories have been developed to analyze visualization quality or effectiveness, variously seeking to explain specific phenomena or enable particular applications.
Here we review a set of evaluatory theories so as to inform our choice to focus on AVD.

Some theories metricize quality, which enables the use of computational measurement as a way to automatically evaluate visualizations.
Behrisch \etal \cite{behrisch2018quality} describe the state of the art for metric-based qualitative evaluation of visualizations, however each of these metrics is associated with particular visual forms.
Mackinlay~\cite{mackinlay1986automating} describes notions of effectiveness and expressiveness which informs the design of some recommendation systems \cite{lee_insight_2020}.
However, over-reliance on these notions can impede richer designs \cite{bertini2020shouldn}.
Chen \etals \cite{chen2010information, chen2015may} work on using Information Theory to reason about visualization quality yields certain desirable properties---such as generating Shneiderman's mantra\cite{shneiderman1996eyes} as an emergent property. Yet, these techniques are intractable to unassisted humans.
Demiralp \etal\cite{demiralp2014visual} describe a visual embedding-based assertion system, however it requires substantial experimental data tuned to a particular chart.
While metric-based analyses can yield useful insights, they can be non-trivial to deploy, are often tuned to specific chart types (forgoing novel forms), and can lack clear interpretation.
We focus on AVD because its assertions are interpretable and can be evaluated without specialized software (i.e. it is human operable).

At the other end of the human-computer agency spectrum are theories that provide human-centered evaluation tools based on heuristic or critical under-pinnings.
Some of these are predicated on personal or philosophical reflection \cite{dork2013critical, bruggemann2020Fold,d2016feminist, vickers2012understanding, bares2020using} which provide useful means for prompting the design process,
but do not include testable assertions by which to judge quality.
Adar and Lee \cite{adar2020communicative} utilize learning objectives as a way to evaluate communicative visualizations, which while providing a definite heuristic by which to evaluate, does not help the designer actually conduct that evaluation.
Wall \etal \cite{wall2018heuristic} build a set of evaluatory heuristics, however their approach requires several domain-experts and that task be pre-established.
Zuk and Carpendale\cite{zuk2006theoretical} conduct heuristic analyses of uncertainty visualizations through the works of Bertin, Tufte, and Ware.
While potentially informative, these theories are piecemeal: they do not provide a holistic description of quality.
Our aims are aligned with these human-operable theories in that we wish to furnish analysts with easy-to-apply tools for analyzing visualizations.
AVD circumvents these issues by providing a systematic framework made up of concrete assertions whose structure helps mitigate potentially biased analyses.

\subsection{Algebraic Visualization Design}\label{sec:avd-background}

At the core of our analysis is Kindlmann and Scheidegger's Algebraic Visualization Design\cite{kindlmann2014algebraic} (AVD).
AVD is a framework for reasoning about the design of data visualizations through their intrinsic symmetries.
Similar to how one may understand the properties of a triangle by identifying which rotations and reflections yield symmetries or asymmetries,
AVD understands a visualization by exploring the effect of changes in data (referred to as $\alpha$s) and the corresponding changes in the resulting image ($\omega$s).
This is mechanized by asserting that every $\alpha$ should have a corresponding $\omega$, that is, these changes should commute:
\begin{figure}[h!] 
    \vspace{-0.1in}
    \centering
    \includegraphics[width=\columnwidth]{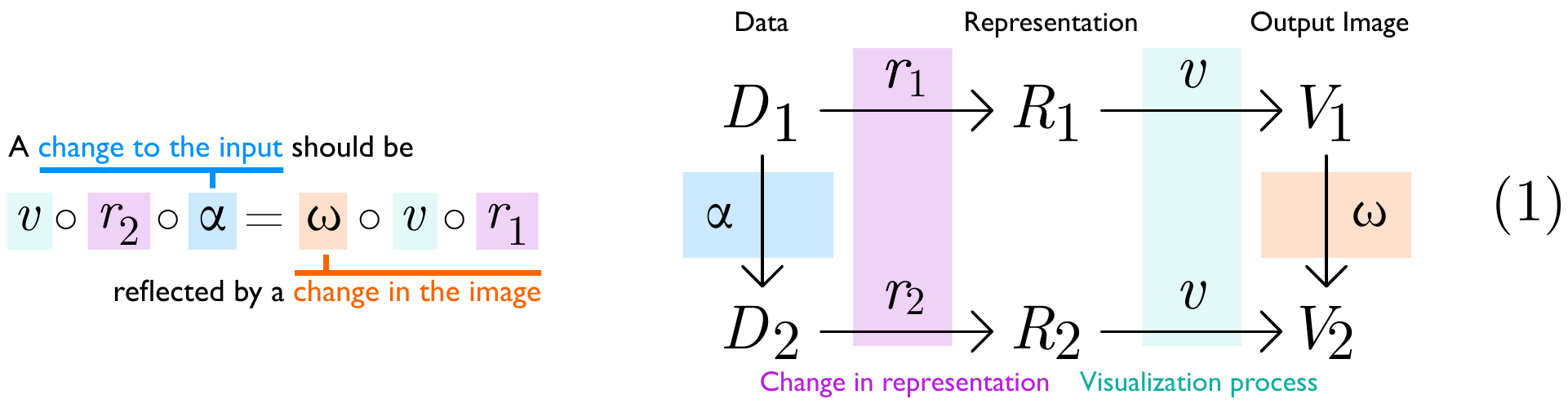}
    \vspace{-0.3in}
\end{figure}

\hallucs{} and \confusers{}, two important failure modes (or asymmetries), occur when this commutativity assertion is not fulfilled (\figref{fig:avd-error-modes}).
If the image changes significantly  as a result of only a small or superficial data change, then it has a \halluc{}:
non-data is depicted in a way that risks appearing meaningful.
Merely reordering columns of a radar chart can dramatically change its enclosed area and visual appearance, a \halluc{}.
If the image does not appear to change with a significant data change, then it has a \confuser{}: a way in which the viewer will be blind to the data.
Standard summary statistics (e.g. mean and variance) for Anscombe's quartet~\cite{anscomb} yield \confusers{} across column changes ($\alpha$s).
Visualizations are probed for these states by adversarially selecting $\alpha$s and $\omega$s that might surface them.

All visualizations have \hallucs{} and \confusers{}. Just as no chart is perfect for all occasions, no graphic is a panacea under algebraic analysis. Instead, \confusers{} must be chosen according to task, and \hallucs{} generally minimized.
These asymmetries help identify properties of the visualizations under consideration in a manner that is only minimally reliant on the observer's experience;
thereby providing distance from the human responses which might otherwise cloud an analysis of a novel graphic.

\begin{figure}[t] 
    \centering
    \includegraphics[width=\columnwidth]{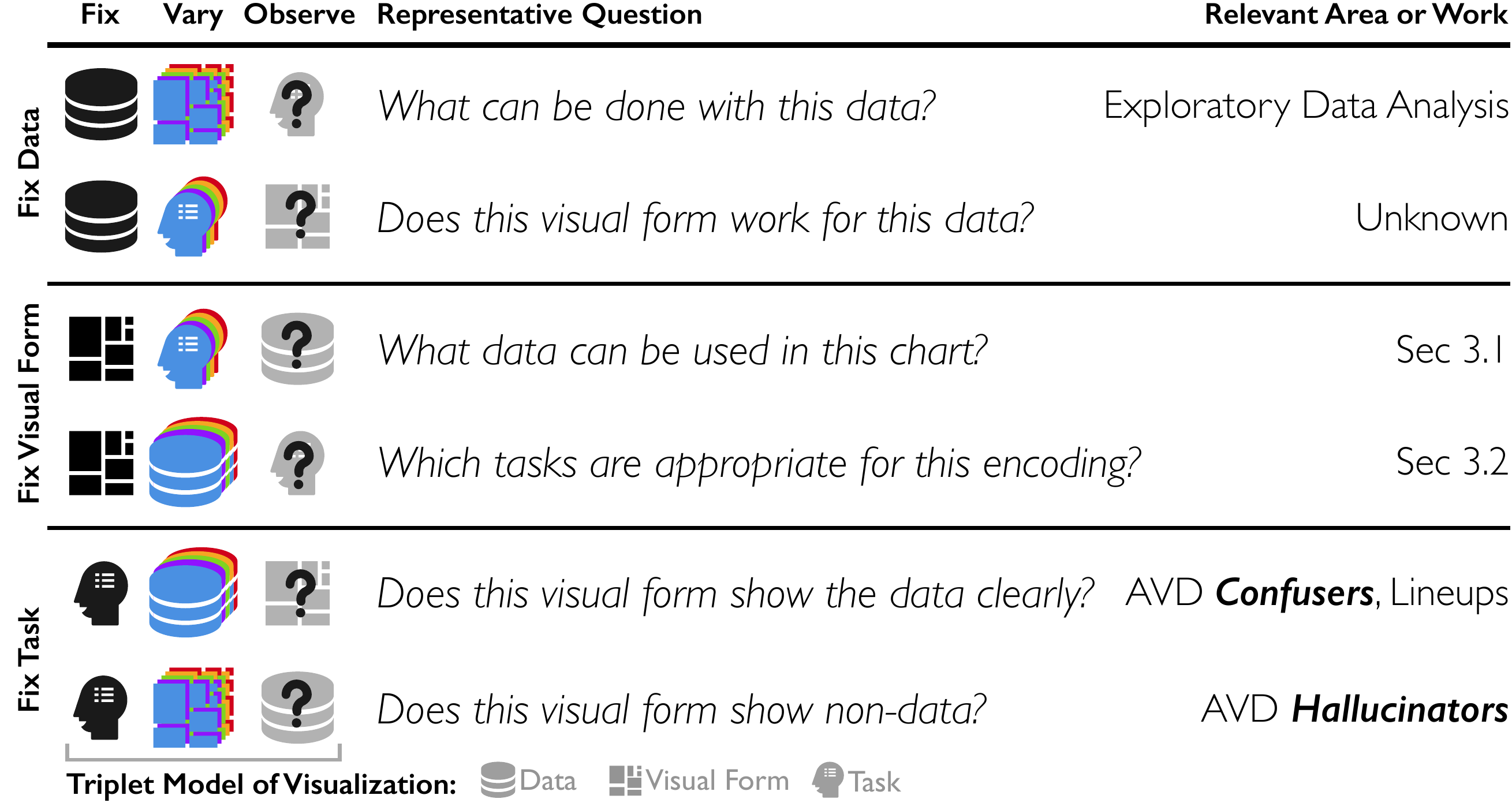}
    \caption{
        Many questions can be answered about visualization quality by taking a variational view of Hu \etals{}\cite{hu2019viznet} triplet model.
        We focus on one of the many analysis families in this formulation by deriving properties of \taco{}s by varying task and data.
    }
    \label{fig:methodology-fig}
    \spaceit
\end{figure}

\section{Algebraic Visualization Analysis of the Table Cartogram}\label{section:analysis}

We now carry out our algebraic analysis of the table cartogram.
We organize our discussion by noting, per Hu \etal \cite{hu2019viznet}, that every visualization can be viewed as a triplet of (data, visual form, task).
Only by correctly matching all three components will a particular visualization have value.
A dataset and visual form might be well matched, but if their combination lacks details necessary to complete the task, then the resulting graphic will be ineffective.

As we are interested in a particular chart form (\taco{}s), we are left with two elements of this triplet to consider: data and task.
Thus, we form our study around two corresponding questions: \emph{What data can be used in this chart?} (\secref{sec:data-analysis}) and \emph{Which tasks are appropriate?} (\secref{sec:task-analysis}).
These questions capture many possible usage concerns, informing when and how \taco{}s might be  used---although, as we discuss,
there are properties that cannot be understood with this approach.

\parahead{Methods}
We perform this analysis by using a reframing of AVD.
In past work AVD has usually been employed to guide the design process, wherein designs are invalidated for a fixed task by variation of data\cite{kindlmann2014algebraic, wood2018design}.
We invert this procedure by using AVD as a way to guide data selection and task design for a given visual form, which we refer to as \emph{Algebraic Visualization Analysis}.

We illustrate this variational scheme in \figref{fig:methodology-fig}.
This locates our work among both that of AVD as well as Exploratory Data Analysis (which fits into this framing by viewing it as a process in which data is fixed and encoding is varied to see what tasks can be fulfilled).
Wickham \etals \cite{wickham2010graphical} lineup protocol is closely related to past usage of AVD, fixing task and varying data to invalidate a design (though it can only be used once per dataset).
Just as in the original AVD study, we select adversarial $\alpha$s and $\omega$s, potentially yielding one of the states in \figref{fig:avd-error-modes}. We diverge from past usage by employing these methods not as a design tool, but as a way to better understand a chart form.
Not all questions need to be answered by explicit variation: some failure modes can be identified by simply applying definitions.
We forgo questions of interactivity as AVD is unable to reason about it, which, while unfortunate for real world applicability, usefully limits the scope of our discussion.

\parahead{Warm up: A Hallucinator}
As a warm up, we consider a prominent \taco{} \halluc{}.
The under-constrained nature of the \taco{} allows there to be multiple equally accurate \taco{} layouts from the same data. \figref{fig:apples-to-apples} shows an example, garnered from changing algorithms, however significant variance can result from seemingly-innocuous parameter changes \cite{mcnutt2020Minimally}.
This multiplicity is a \halluc{}.
While selecting \figref{fig:apples-to-apples}(b) over \figref{fig:apples-to-apples}(a) may be perceptually motivated (as it is easier to compare rectangular shapes), the value of selecting a particular starting condition in an optimization algorithm can be unpredictable and ambiguous.
This contrasts tree maps, where algorithm design is well understood, predictable, and usually has meaningful design implications \cite{scheibel2020survey}.

One way to address this failure mode is to impose additional constraints---such as by minimizing bearing angle differences---which causes there to be a single \emph{``correct''} layout\cite{inoue2020optimization}.
While these criteria can yield more rectangular displays, their selection is arbitrary and an artifact only of designers' preferences, an ambiguity which may in turn hold another \halluc{}.
A reader comparing \taco{}s for different datasets prepared according to different heuristics (but who is unaware of this selection) may be deceived.

This flexibility is both a blight and a boon.
The multiplicity of outputs creates space for uncertainty: \textit{how can the reader know their interpretation of the chart is correct?}
Yet this same property  offers a great measure of freedom to chart designers to create visually interesting effects.
The tension found between designer freedom and potential reader mistrust suggests that \taco{}s should not be used in decision-making contexts.
Just as geographic cartograms are typically used to give big-picture summaries\cite{nusrat2016state} rather than in data analysis,
we argue that \taco{}s are best applied in situations where the task involves the readers awareness of the presentation medium itself (in what might be called \emph{autotelic visualizations}) or in casual consumption contexts (such as in \emph{enjoy} tasks\cite{brehmer2013multi}).

\begin{figure}[t] 
      \centering
      \includegraphics[width=\columnwidth]{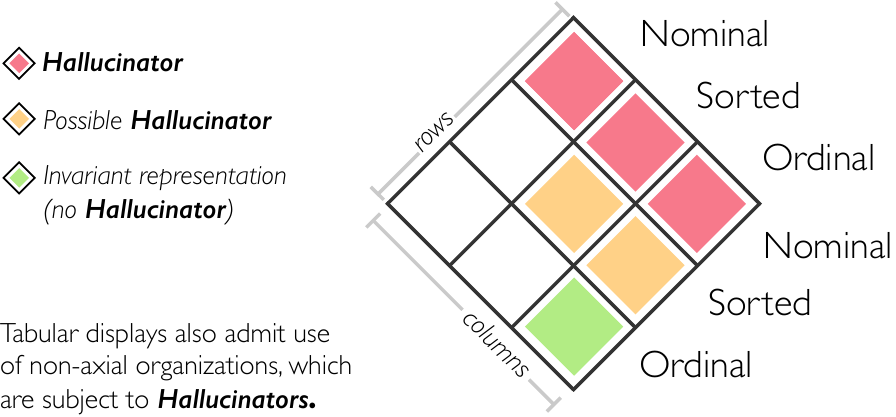}
      \caption{
            The tabular form of \taco{} input data gives way to seven transpose unique axial data types.
      }
      \label{fig:data-space}
      \spaceit
\end{figure}

\subsection{What Data Can Be Used?}\label{sec:data-analysis}

Understanding \taco{}s' potential utility starts with identifying their valid data inputs. Here we begin our discussion in earnest by analyzing the space of possible data types, values, and sizes.

\subsubsection{Data Type}

We first focus on understanding the types of inputs that can be appropriately visualized by \taco{}s. Definitionally a \taco{} can only be meaningfully computed on a 2D-table of scalars. Yet this still leaves a large space of potentially allowable tables.

We argue that the scalars that make up such tables must share a single unit of measurement.
Inputs with heterogeneous units would multiply define the meaning of area, and hence undermine the interpretability of the output.
In particular, by changing the relative definition of the potentially unrelated units, arbitrary changes to the visualization would be induced; a \halluc{}.
A similar \halluc{} is found in dual-y-axis charts\cite{mcnutt2020surfacing},
which correlate unrelated units by an arbitrary choice of normalization \cite{dual-y-axis}.

\begin{figure}[t] 
      \centering
      \includegraphics[width=\columnwidth]{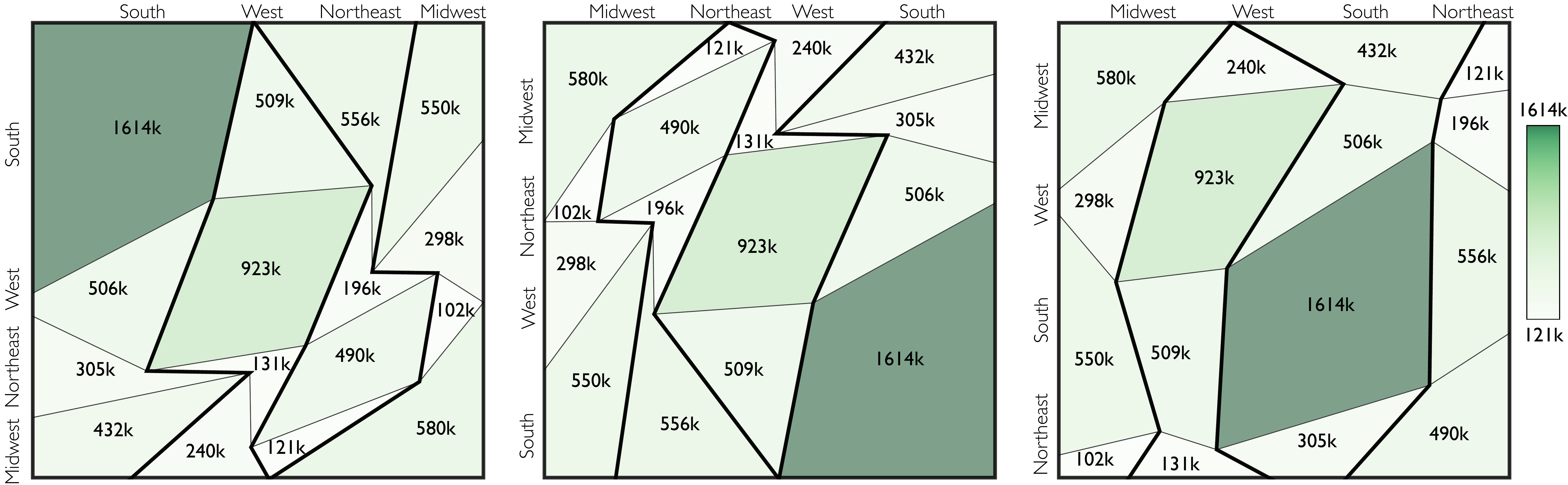}
      \caption{
            \taco{}s intermingle data with layout. Here 2016 US regional migration~\cite{census-state-to-state} is permuted
            across axial orderings.
            Each ordering gives the bolded columns visually different layouts, yet these transformations ($\alpha$s) are not meaningful: a \halluc{}.
      }
      \label{fig:region-to-region-compare}
      \spaceit
\end{figure}

Next, to better understand this remaining space of inputs we organize it into a typology defined by row and column data (\figref{fig:data-space}).
We form this space by taking all pairwise combinations of elements in a slight expansion of Stevens' discrete types\cite{stevens1946theory}: nominal, sorted nominal (per Inoue and Li \cite{inoue2020optimization}), and ordinal.
Beyond these simple types, tables afford a large space of hierarchical organizations~\cite{hurst2006towards, bakke2013automatic}, such as nested-pivot tables.
We simplify these higher-order types by classifying them as nominal or ordinal by whether their hierarchical order is nominal or ordinal.
This yields six transpose unique axial combinations.
Finally there are non-axial orderings, which use a layout algorithm to mold non-tabular data into a grid, such as in a waffle plot (see Appendix).
While this case merits further study, we focus on the more common case of tables whose meaning is described by their axes.

With this model in mind, we argue that tables featuring either an intrinsic bidirectional ordering (as in calendars) or those that are sorted are preferable to those with nominal axes. In \figref{fig:region-to-region-compare} we show region to region migration in the US across row and column permutations (AVD $\alpha$s), each of which have equal accuracy.
Each ordering is equally valid, as this data does not have an intrinsic order.
However, the visual form of the cells and highlighted columns is inconsistent across these options which reveals that this data type is a \halluc{}.
This implies that only tables with non-nominal orderings should be used. This excludes, for instance, data found in CSVs or tidy tables whose row order is usually intended to be non-meaningful, as well as data that can be coherently expressed under multiple table projections
(such as tidyr's \verb+pivot_long+ and \verb+pivot_wide+ \cite{tidyverse}).
Further, we suggest that flat tables should be preferred as tabular hierarchies are often non-ordinal.

Slingsby \etal \cite{slingsby2009configuring} describe a set of guidelines for organizing hierarchical datasets in tree maps, arguing that nominal data should be treated with a consistent ordering to facilitate legibility.
This agrees with our assertions that non-nominal data is preferable, as well Inoue and Li's \cite{inoue2020optimization} notion that sorted-by-similarity axes are more effective for \taco{}s.
To this latter point: the validity of sorted nominals is complicated by the arbitrariness of choice of sorting algorithm
---while possibly motivated by the maintenance of particular metrics or aesthetics,
may yield a \halluc{}.

\begin{figure}[t] 
    \centering
    \includegraphics[width=\columnwidth]{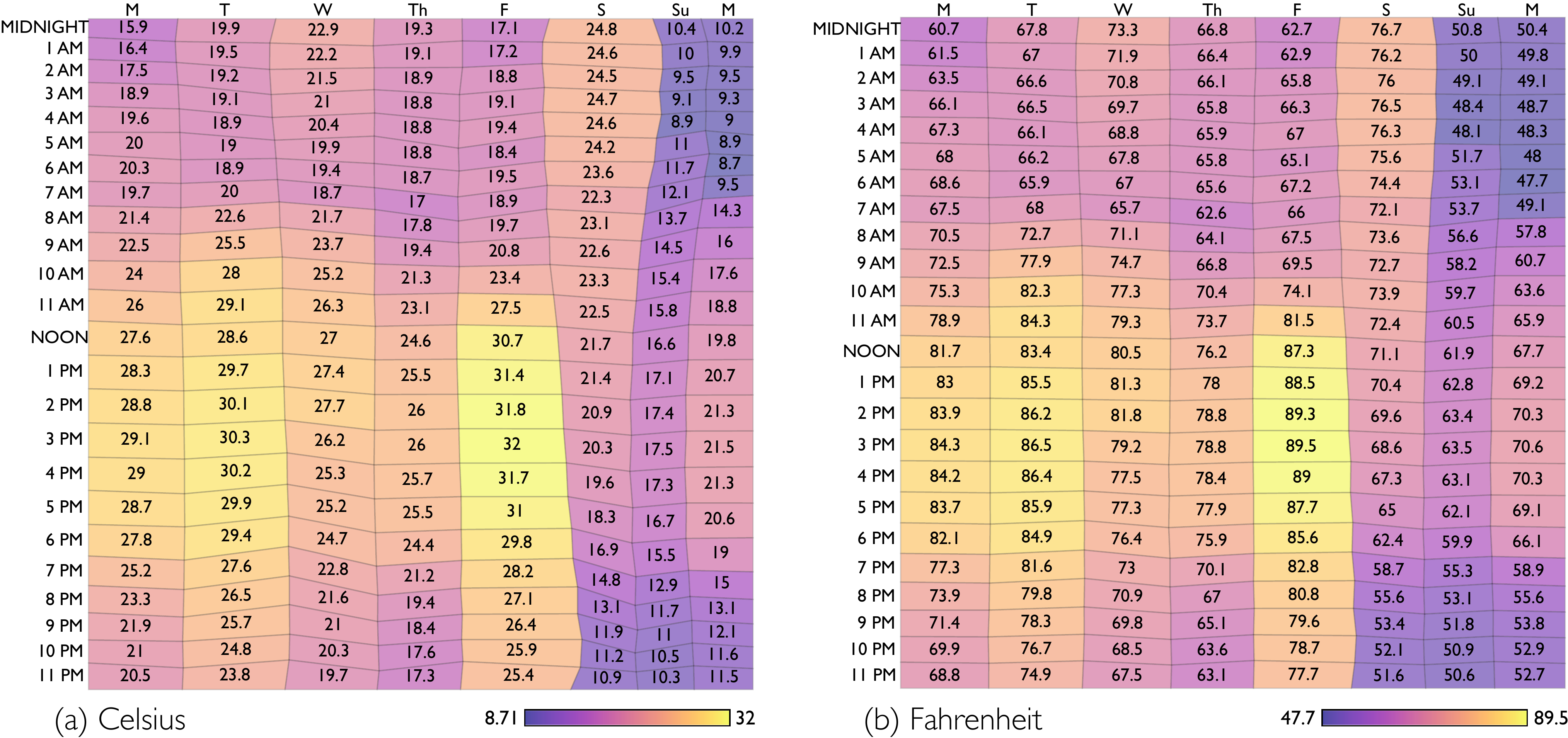}
    \caption{
        Temperature at O'Hare Airport Sept. 16-23, 2018
        \cite{ohare-weather}.
        Despite the underlying data equality, these \taco{}s are visually different: (a) shows the changes more dramatically.
        This is a \halluc{}, but can be tamed to serve particular tasks.
    }
    \label{fig:temp}
\end{figure}

\subsubsection{Data Value}\label{sec:data-value}

Next, we continue to limit the space of possible inputs by considering appropriate cell values.
A natural question to ask is whether \taco{}s support cell values beyond the positive scalars ascribed to them in our definition.
Consider an $\alpha$ in which a particular value is negated.
Our definition of \taco{}s does not provide meaning for a negative area, and as such this is undefined behavior.
The result will be implementation dependent, perhaps being treated as a positive value (yielding a \confuser{}) or collapsing the planar topology.
Next, consider an $\alpha$ in which a value is set to zero.
The collapsed grid topology would then present non-adjacent cells next to each other.
These  collapses break our premise that we are keeping the visual form fixed.
This leaves us with positive scalars.

Within positive scalars there are two possible Stevens' data types\cite{stevens1946theory}: ratio and interval.
Area encoding has a natural root of zero (zero area indicating zero value), which pairs well with ratio encodings (which are defined by having a meaningful zero).
The \taco{}s pliability to interval data on the other hand is less clear. Data for which an affine transform, such as $\hat{x} \rightarrow m x + b$, is merely a re-representation (such as changing between Celsius and Fahrenheit) belongs to the \emph{interval} scale.
That is, interval data are defined by having a zero whose meaning is not intrinsic to the type.

The incongruity between interval data's non-meaningful zero and \taco{}s' rooted zero would seem to indicate that \taco{}s should not be used with interval data.
Yet, we suggest that a \taco{} of interval data can nonetheless be informative, depending on the task.
\figref{fig:temp} shows a set of temperatures as \taco{}s in different units.
The differences between these charts is due to the (arbitrary) choice of unit selection.
For interval data, this choice can yield a \halluc{} (if units are chosen to overly magnify differences in value) or a \confuser{}
(as might be the case if Kelvin had been represented as well).
Yet, what could be considered a critical failure mode for \taco{}s can also be wielded as an intentional design choice.
\emph{If the goal is to detect extrema or trends}, the Celsius units may be a better design choice, because this data is mapped to a wider variety of areas, even though there is nothing essential about areas being proportional to degrees Celsius.
More extreme variations (to sub-freezing temperatures) would require a different affine transform to map the values to legible area variations, analogous to intentionally setting the axes bounds of scatter plots in a data-dependent way.

\begin{figure}[t] 
    \centering
    \includegraphics[width=\columnwidth]{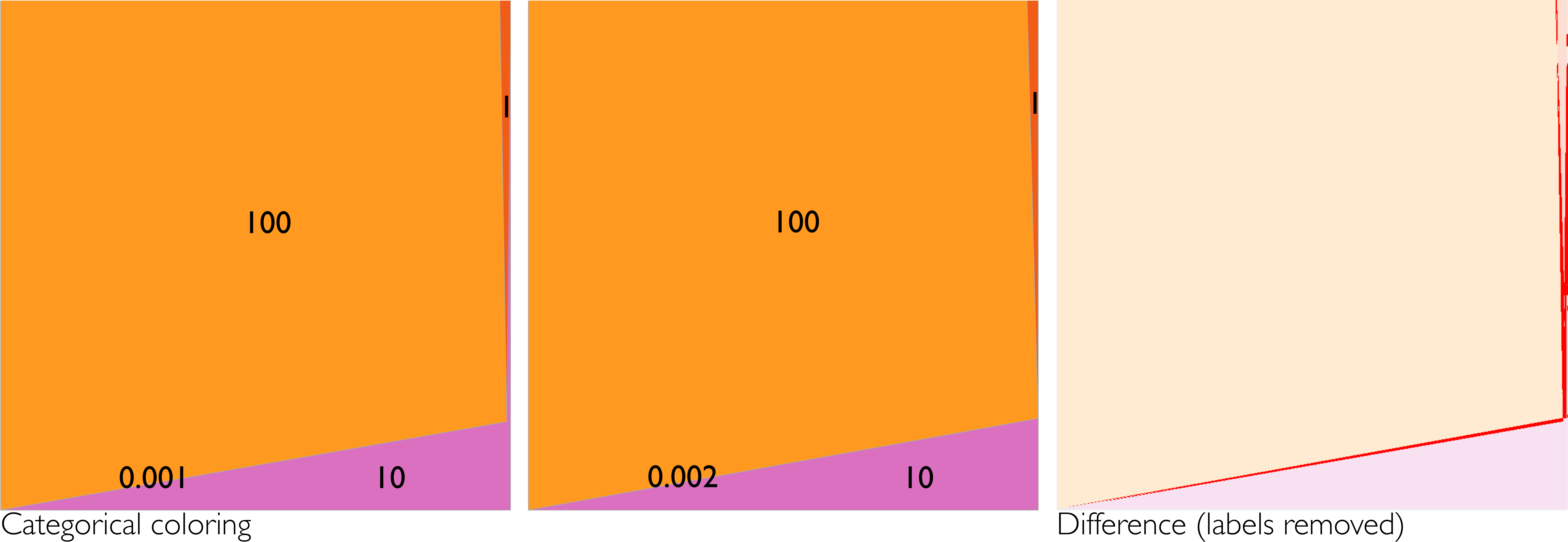}
    \caption{
        \taco{}s containing data ranges with magnitudes of $10^{5}$. Despite doubling of the smallest value in this toy dataset
        (an $\alpha$),
        the images appear nearly identical: a \confuser{}.
        The visual subtly of the difference might cause it to be overlooked.
    }
    \label{fig:min-size-compare}
\end{figure}

\begin{figure*}
      \includegraphics[width=\linewidth]{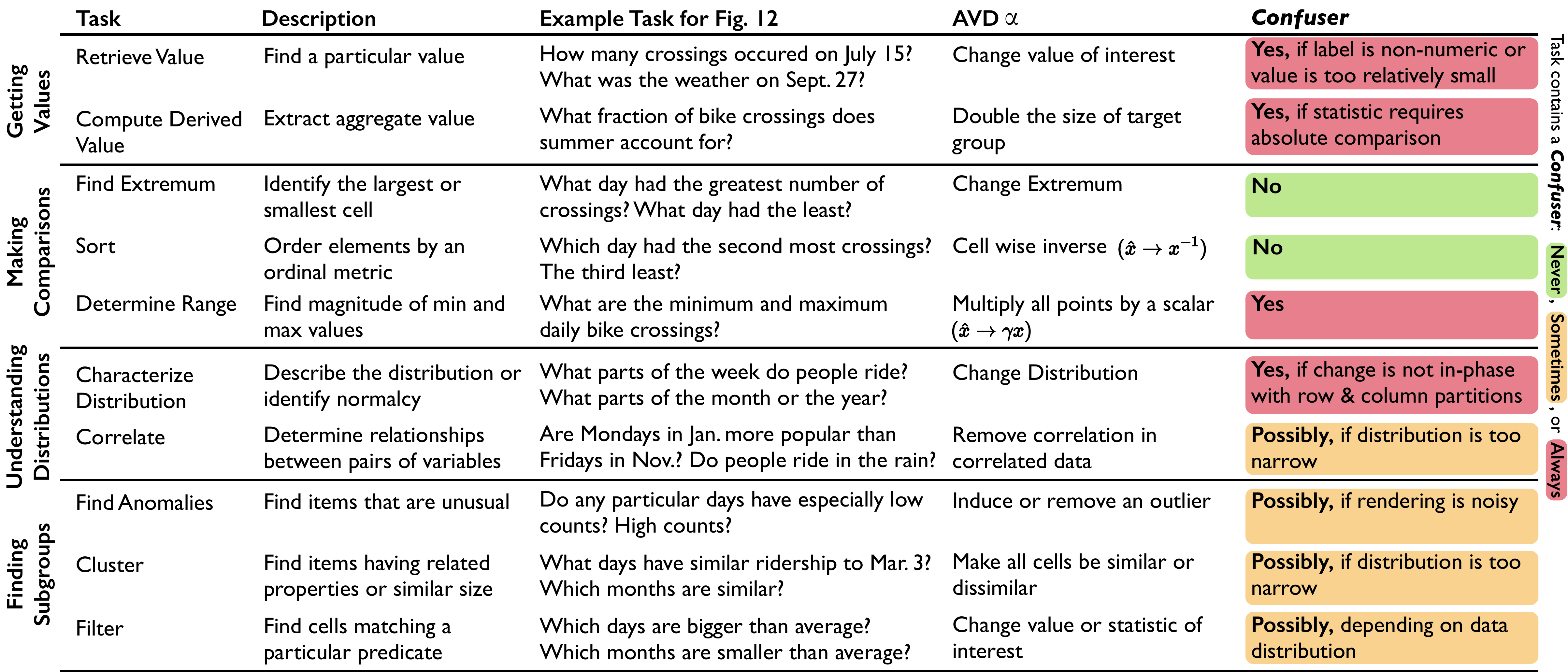}
      \vspace{-0.2in}
      \caption{
            \label{fig:long-table}
            The \taco{}'s performance on a low-level task typology\cite{amar2005low} guided by whether those tasks exhibit \confusers{} for relevant $\alpha$s.
      }
      \spaceit
\end{figure*}

\subsubsection{Data Size}
All visualizations have limitations in the volume and range of values they can support while remaining comprehensible \cite{munzner2014visualization}.
Many guidelines place limits on these ranges, however there are typically exceptions or special cases to such bounds.
We suggest that broadly applicable guidelines are defined with a measure of malleability or ambiguity.
In this vein we  construct a loose bound on appropriate data range and size.

We first look at range. Consider a table of values $x_{ij}\in\mathbb{R}^+$ with sum
$\sigma=\sum_{ij}x_{ij}$
. We induce a normalizing transform
$\hat{x}_{ij}=x_{ij} \sigma^{-1}$.
The area of a cell with no error is
$a_{ij}=whx_{ij}\sigma^{-1}=wh\hat{x}_{ij}$,
where $w$ and $h$ are the graphic's height and width.
We select a minimum legible area for a cell to be 1 pixel, as sizes below this are hypothetically not-representable in a pixel based system.
Other reasonable values might be informed by notions such as area JND \cite{raidvee2020perception}, however this simplistic choice serves our simplistic goals.
We identify the minimum value, $\min{x_{ij}}=x_{min}$, for which changes would \emph{not be invisible} (all $\alpha$s yield a \confuser{}).
\begin{equation}\tag{2}
    a_{min} = 1 = \hat{x}_{min}wh
    \hspace{0.3in}
    \Rightarrow
    \hspace{0.3in}
    \hat{x}_{min} = (wh)^{-1}
\end{equation}
For a chart with $w=500$ and $h=500$---as many \taco{}s in this paper are---then $\hat{x}_{min} = 2.5 \times 10^{-5}$. That is, \taco{}s with a range wider than $10^{5}$ will possess a \confuser{} as any changes to its smallest values will be illegible, as in \figref{fig:min-size-compare}. This bound is conservative, as it describes the ratio between min and sum, and not the min and max (which would be tighter).
However, as we sought a rough bound, this is sufficient.
In practice we find that it is usually better to use a range smaller than $10^{3}$, however we do not argue that wider ranges (but $<10^5$) are unsound. Such ranges may be legible if the data has a regular structure (as in Appendix \figref{fig:teaser}) and the task is not dependent on individual values.

Appropriate row and column cardinality is more intertwined with task for \taco{}s than range.
For instance, a table consisting of a small number of rows and columns are preferable when those individual values have relevance to the task under-consideration (such as \task{Retrieve Value}, cf. \secref{sec:get-value}).
Again, larger tables can be shown if the goal relates to aggregate relationships or smoothly varying distributions, since their interpretation involves trends rather than individual values.
A rudimentary assertion would hold that each column or row should possess at least one pixel; suggesting a naive upper bound of $|rows| \leq h$ and $|columns| \leq w$.
Yet, even this simple guideline has holes: adjacent cells of a common color with partial pixel areas can be understood as constituting larger units, allowing comparison of aggregates.

These approximations highlight a weakness in our approach: not every question can be answered by these analytic means.
However, we believe that these approximations are sufficient to aid effective usage.
These guidelines could be more closely examined (and made more precise) through a user study.

\subsection{What Tasks Can Be Used?}\label{sec:task-analysis}

Next we conduct an algebraic task analysis by utilizing Amar \etals task typology \cite{amar2005low}, organized into four themes (\figref{fig:long-table}).
This typology describes a set of low-level functions which users might perform on visualizations.
We focus on this typology (among others\cite{bertini2020shouldn, brehmer2013multi}) because of its simplicity and ubiquity.
We use this general typology---rather than one focused specifically on cartograms\cite{nusratCartogramTask}---because it is domain agnostic, which enables comparison with non-geographic charts.

\figref{fig:long-table} summarizes this analysis, categorizing each task by whether a relevant $\alpha$ yields a \confuser{}.
We focus here on \confusers{} as they are more closely related to task than \hallucs{}\cite{KindlmannAlgebraicVisPedagogyPDV2016}.
To ground this analysis we include a specific task for \figref{fig:seattle-bikes}, which is discussed in greater length in \secref{sec:case-study}.

\subsubsection{Getting Values}\label{sec:get-value}
\textbf{Tasks}: \task{Retrieve Value}, \task{Compute Derived Value}\\
\textbf{Discussion}:
\taco{}s, like unadjusted tables, appear to be well-suited to the \task{Retrieve Value}, but can exhibit a \confuser{} if not labeled appropriately or if the relevant value is too small to visually resolve.
The \taco{}'s tabular structure allows looking values up by rows and columns, even in the presence of distortion---although this may be impeded by layouts that are highly dissimilar to grids. Changes to these values might be invisible (yielding a \confuser{}) if the change is small or the cells are not labeled with their corresponding value.
This suggests that, like both geographic cartograms\cite{nusrat2016state} and tree maps\cite{friendly1994mosaic}, \taco{}s are more effective when used in conjunction with a secondary visual encoding, such as color or text, as it facilitates easier value retrieval than simply using area alone \cite{inoue2020optimization}.
We argue that, given the low accuracy of the perceptual system for understanding numeric values through area \cite{mackinlay1986automating}, that value retrieval hinges on the presence of secondary annotations, and possesses a \confuser{} otherwise.
Beyond identifying their necessity, AVD's coarse-grained assertions offer little insight into the perceptual role that secondary encodings play.
Subsequent work should investigate the effect that secondary encodings have on the completion of rudimentary visualization tasks.

\begin{figure}[t] 
      \centering
      \includegraphics[width=\columnwidth]{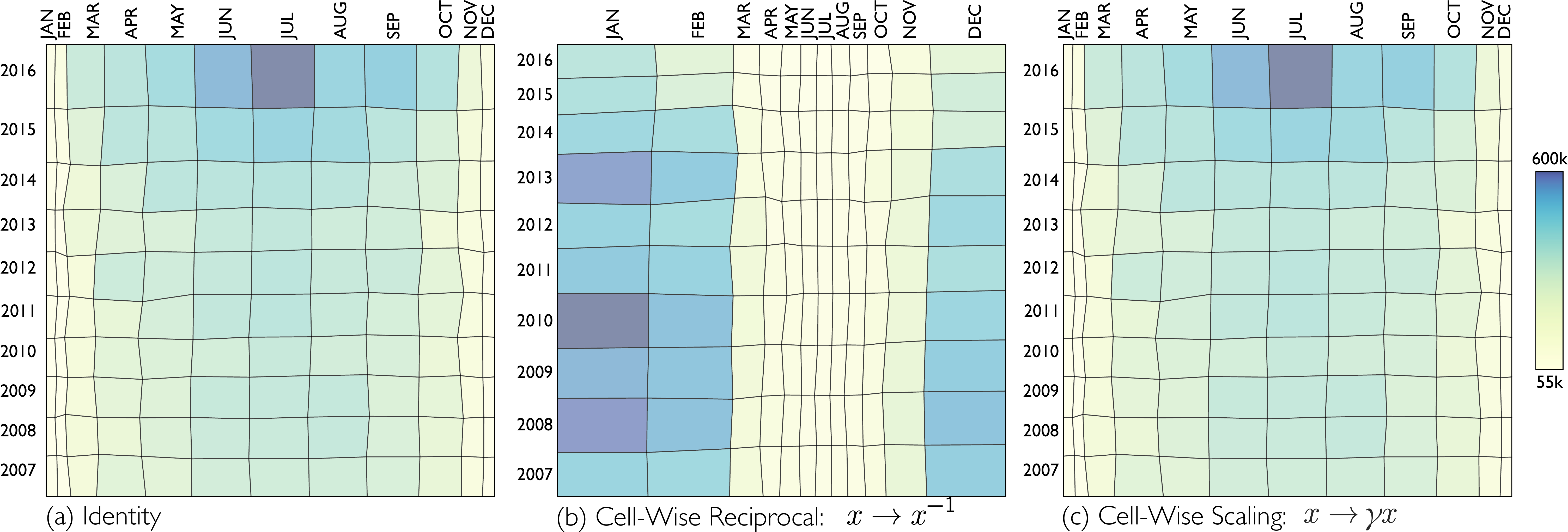}
      \caption{
            Visitorship to Zion National Park
            \protect\cite{zion-stats}
            across $\alpha$s.
            (b) demonstrates that \taco{}s accurately maintain data order across an order inverting transformation, while (c) highlights a \confuser{}: changes to scale are invisible.
      }
      \label{fig:zion-with-recip}
      \spaceit
\end{figure}

Less dependent on the specifics of secondary encodings is \task{Compute Derived Value} which involves the visual addition of cells to form an aggregate value.
There are a variety of statistics that may be evaluated, however only those involving \emph{relative} values are supported by \taco{}s, as those related to \emph{absolute} magnitudes will possess a \confuser{} (per \figref{fig:zion-with-recip}c rescalings are invisible).
This preference for relative comparison in conjunction with the availability of visual addition suggests that they support part-to-whole and part-to-part relationships, just as in pie charts.
We suggest that---also similarly to pie charts\cite{kosara-pie-charts}---\taco{}s only support the visual summation of adjacent parts,
however this is beyond the scope of AVD's simple assertions, so we do not verify it.

\subsubsection{Making Comparisons} \label{sec:compare}
\textbf{Tasks}: \task{Find Extremum}, \task{Sort}, \task{Determine Range}\\
\textbf{Discussion}: As we have seen, areas in \taco{}s are proportional to data values \emph{relative} to the sum of all values. This enables some types of comparisons---although those dependent on non-relative comparison exhibit \confusers{}.
The relative nature of area comparison enables identification of a consistent visual ordering of the data values, as it is typically easy to identify which of two shapes is larger \cite{Cleveland-GraphPercept-1984}, though the irregular shapes of some cells can affect this ordering,
as can small differences in areas \cite{martinez1973ranking, krider2001pizzas}.
It is notable that area perception of blobs (as these quadrilaterals might be interpreted) yield more accurate comparison than area comparisons in circles \cite{cleveland1986experiment},
which suggests that \taco{}s may be preferable to those encodings in some contexts.

When per-cell data values are replaced by their reciprocals (an $\alpha$) all relative orderings are reversed in a consistent way, as in \figref{fig:zion-with-recip}(b).
This suggests that \taco{}s are consonant with ordinal comparisons as the rendered visual ordering of the entire visualization is consistent across transformation, and thus pliable to  \task{Sort} and \task{Find Extremum}, which both rely on relative comparisons (e.g which of two cells is bigger).
Yet, not all comparisons are supported.  \figref{fig:zion-with-recip} (c) shows that a uniform scaling (another $\alpha$) is invisible, and hence a \confuser{} for \task{Determine Range}, which asks viewers to find  absolute values of extrema. This is consistent with familiar graphics such as tree maps or pie charts, which have \confusers{} for tasks depending on \emph{absolute} scale \cite{kindlmann2014algebraic}.

\begin{figure}[t]
    \centering
    \includegraphics[width=\linewidth]{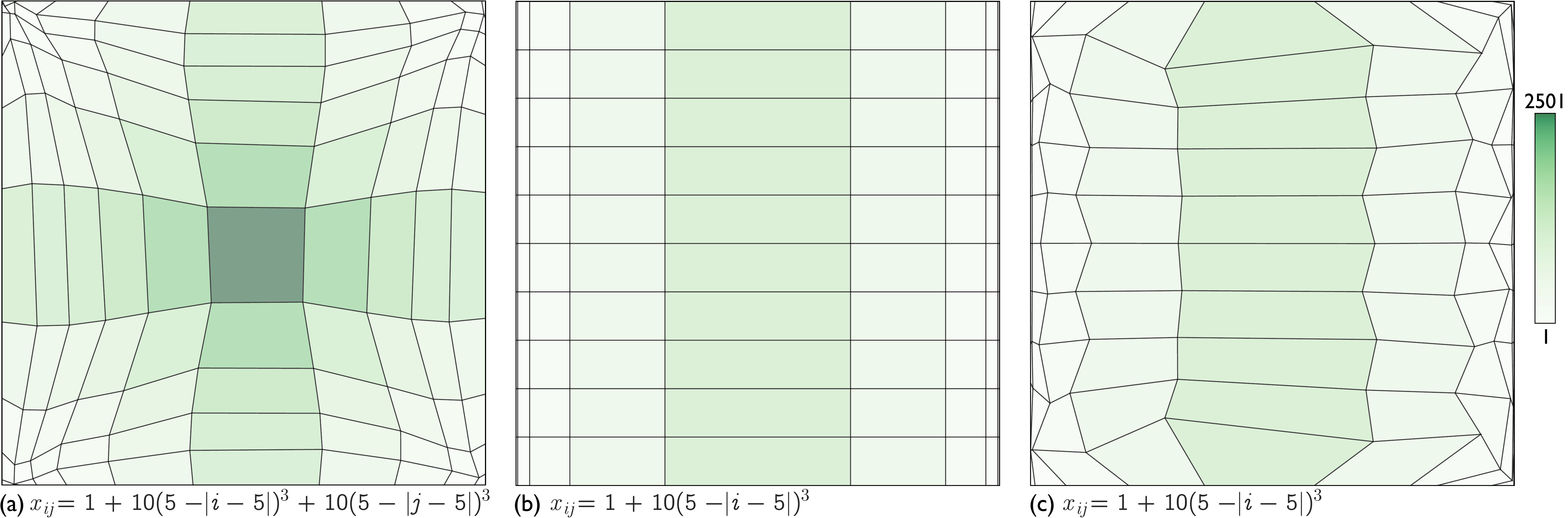}
    \caption{
        Strong correlations between axes can be apparent, however it may be impaired by the multi-layout \halluc{}:
        which can suggest non-existent correlations (as in (c)'s top and bottom rows). (a)'s axes are  correlated, while in (b) and (c) that correlation is removed (an $\alpha$) by making the x distribution uniform.
    }
    \label{fig:correlation-compare}

\end{figure}

\subsubsection{Understanding Distributions}\label{sec:distributions}
\textbf{Tasks}: \task{Characterize Distribution}, \task{Correlate}\\
\textbf{Discussion}:
Displays which privilege individual values
(e.g. cells in \taco{}s)
prompt questions about those value's distribution.
Despite their un-aggregated form \taco{}s may have little value in this context, as they may exhibit \confusers{} (depending on distribution).

We can probe \task{Characterize Distribution} with an $\alpha$ that changes the distribution. Changes will only be visible if they are in-phase with the partitions of the rows and columns. For instance, shifting all visits to a park (such as in \figref{fig:zion-with-recip}) to happen at night will be invisible on a month calendar, but changing the number of visitors on a particular day (such as by inducing a holiday) will be visible.
For interval data, \task{Characterize Distribution} can yield a \halluc{} or a \confuser{}, depending on the selection of units (\secref{sec:data-value}).

To investigate \task{Correlate} we can employ an $\alpha$ that removes an extant correlation, such as by shifting to a uniform distribution, as in \figref{fig:correlation-compare}. Again,
the legibility of correlations is contingent on their relationship being visible under the selected row and column partitions.
This task also involves consideration of correlation magnitude; for which the \taco{} is poorly tuned,
as it's variable layouts can hallucinate weak or non-existent correlations. For instance, the top and bottom rows of \figref{fig:correlation-compare}(c) appear to be marginally larger than the rest of the rows; implying a non-uniform relationship between value and vertical position.
Beyond correlation between axes, \taco{}s can present correlation between size and secondary encoding (such as color). We forgo considering this property as it necessitates more precise tools than are available in AVD.

\begin{figure}[t] 
      \centering
      \includegraphics[width=\columnwidth]{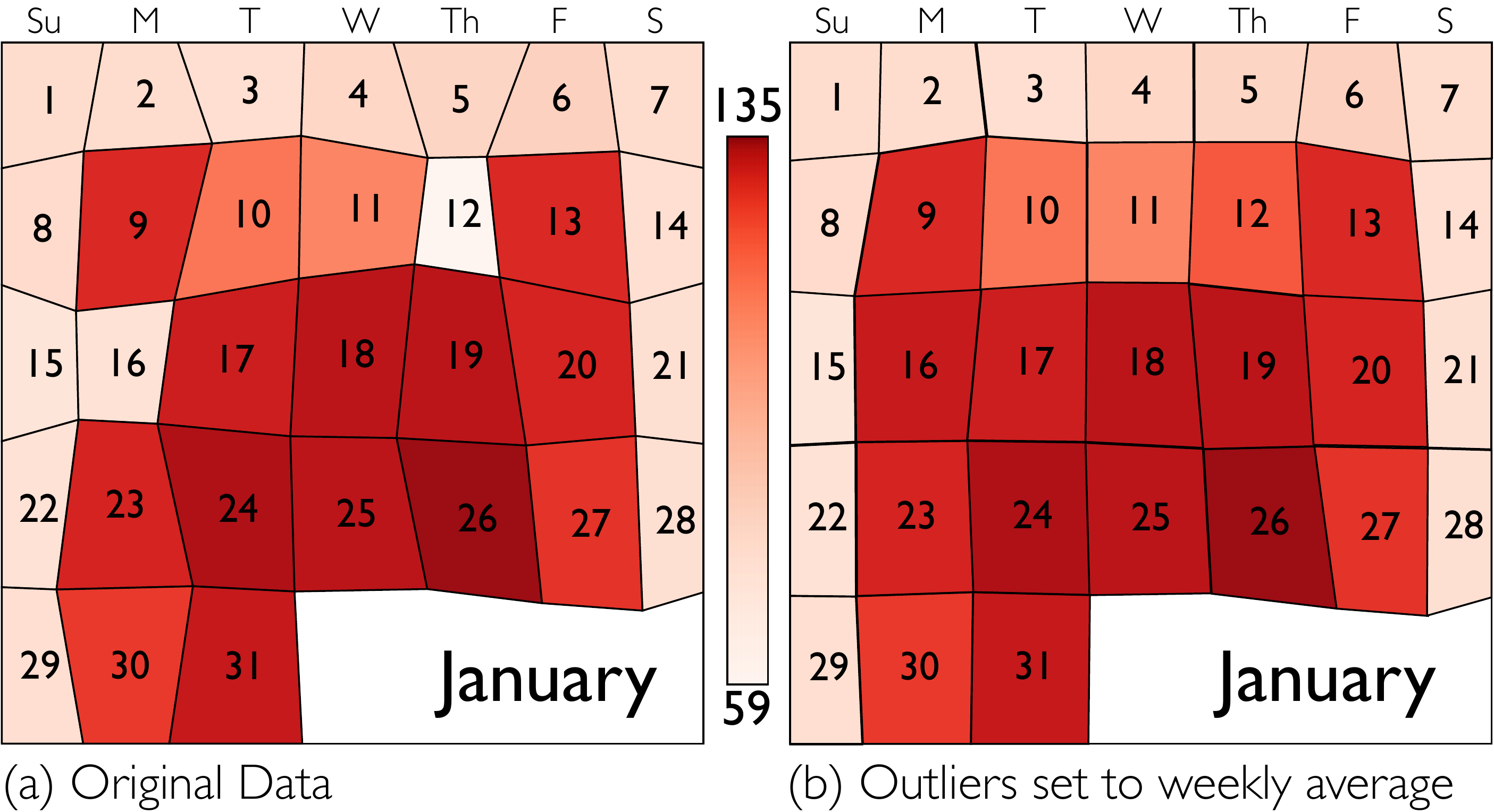}
      \caption{
            \taco{}s highlight outliers by distorting the grid around those cells.
            Here frequency of speed camera violations in Chicago during Jan. 2016~\protect\cite{chicago-red-light} is altered (an $\alpha$) to explore this property.
      }
      \label{fig:cal-diff}
      \spaceit
\end{figure}

\definecolor{alwaysColor}{HTML}{D0021B}
\definecolor{sometimesColor}{HTML}{F5A623}
\definecolor{neverColor}{HTML}{7ED321}

\begin{figure*}[ht!]
    \centering
    \includegraphics[width=\linewidth]{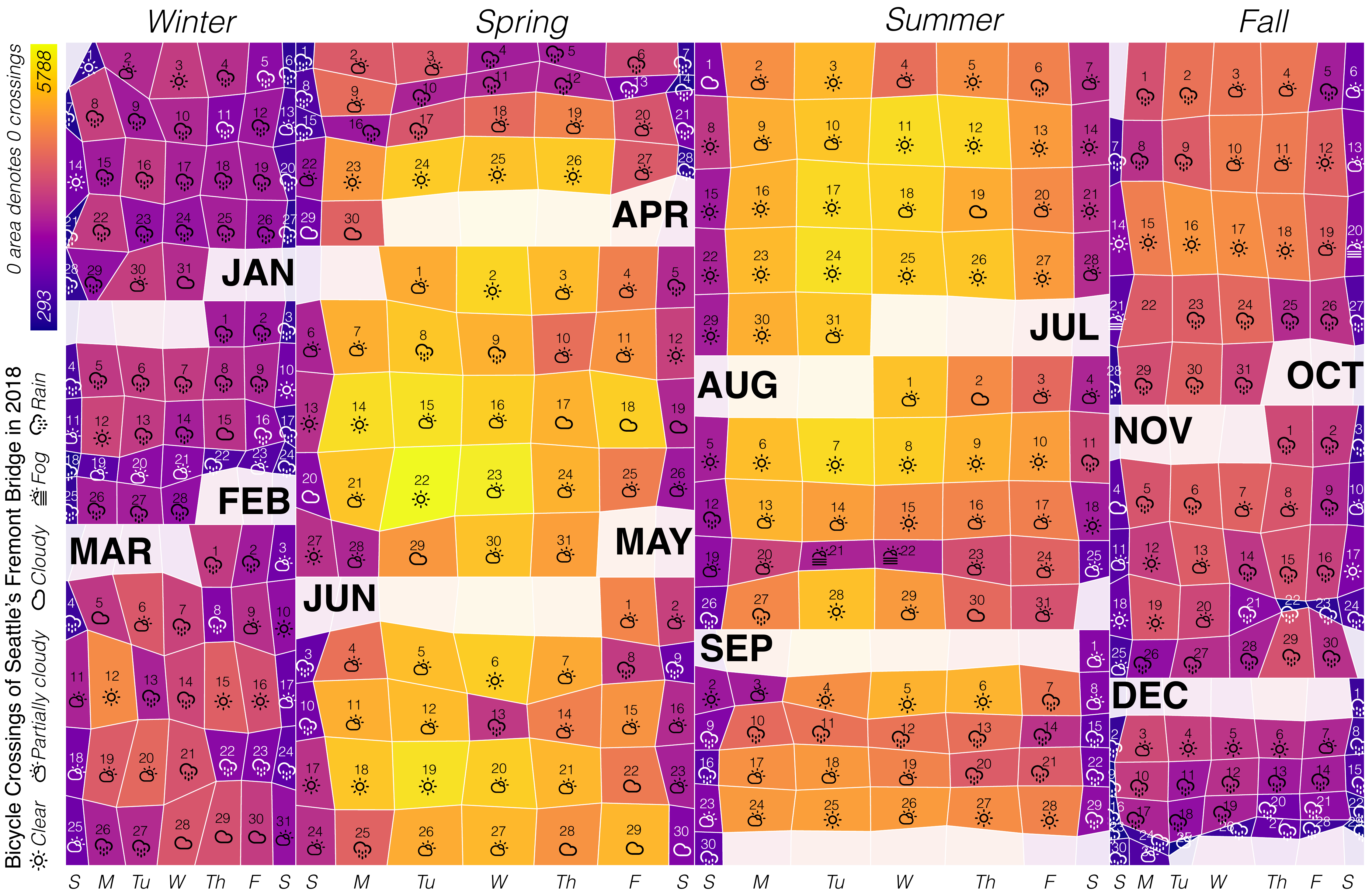}
    \vspace{-0.2in}
    \caption{\label{fig:seattle-bikes}
        Bicycle crossings of the Fremont Bridge in Seattle in 2018 \cite{mcnutt2019Cycles}.
        Months of the year are arranged by quarter in a slice-dice tree map,
        while days of the month are shown as \taco{}s such that each day is sized and colored according to the crossing volume.
    }

    \spaceit

\end{figure*}

\subsubsection{Finding Subgroups}\label{sec:find-anoms}
\textbf{Tasks}: \task{Find Anomalies}, \task{Cluster}, \task{Filter}\\
\textbf{Discussion}:
The \taco{}'s layout forms contiguous paths in rows and columns which can aid investigation of subsets of interest.
The fabric of the table itself is distorted in a coherent manner across contiguous paths for all rows and columns.
We suggest that this distortion facilitates identification of anomalous extrema (\task{Find Anomalies}), while impeding identification of similarly valued cells or groups of cells---as \task{Filter} and \task{Cluster} require.

We can observe \taco{}s handling of \task{Find Anomalies} in \figref{fig:cal-diff}, which removes outliers from smoothly distributed data (an $\alpha$) by setting them to the weekly average.
The visibility of this change indicates \taco{}s are not ineffective for tasks involving identifying outliers.
Those now-average values are more difficult to differentiate from their neighbors.
This can thus impede both \task{Filter} and \task{Cluster} for narrow distributions.
This is in agreement with prior work\cite{martinez1973ranking, krider2001pizzas} which notes that very different comparisons are easier to make than similar ones.

A notable caveat: the \taco{}s' irregularity can cause false positives when trying to detect outliers as cells which may have the same value can be represented as quadrilaterals with greatly differing shapes.
Yet, under an appropriate parameter configuration outliers can be clearly seen  (\figref{fig:cal-diff}).
Unlike  \task{Find Anomalies}, the ease of \task{Filter} or \task{Cluster} seems to be a function of both the separability of subclasses within the distribution, as well as the selection of axis units or partitions. As in other cases, this suggests that there is a \confuser{} when the distribution is narrow, or if the statistic that \task{Filter} is predicated upon possesses a \confuser{}.

\pagebreak

\subsection{Case Study}\label{sec:case-study}

We conclude our analysis with a discussion of a larger example \taco{}\cite{mcnutt2019Cycles}, reproduced in \figref{fig:seattle-bikes}.
We focus on this design because it exemplifies a more complex composition than the  examples we have seen so far, thereby showing more of the \taco{}'s strengths, as well as its weaknesses.
It shows the bicycle crossings on a particular bridge in 2018 annotated with daily weather.
The \taco{} is used here, not just as a graphic unto itself, but as an ingredient in a larger composition, showing each of the months as an independent \taco{} arranged in a slice-dice tree map. Crossing volume is doubly encoded through area and color. The area encoding provides a ratio representation of the value (zero area corresponding to zero crossings), while color shows that quantity through an interval rendering, facilitating ordinal readings---like that yellow cells have a greater number of crossings than others.
This design highlights the \taco{}'s ability to blend with other chart forms, such as tree maps.
In the appendix we discuss another real-world example \cite{pierebean_oc_2020}, which combines \taco{}s and word clouds.

Some tasks are eased by the visual structures of this graphic. As might be expected from our discussion of \task{Find Anomalies}, some outliers are readily visible, such as near American Thanksgiving (Nov. 22), as well as seasonal and monthly trends (which are imparted by the containing tree map).
Some correlations are more readily apparent than others. Weather condition appears to be loosely correlated with increases in ridership (as in the third week of October)---although as we noted in \secref{sec:distributions} magnitude of correlation is hard to judge with \taco{}s.
In contrast, while it is clear that more crossings occur during weekdays than on  weekends, it is completely invisible that more crossings occur during the middle of the day on the weekends, as well as before and after business hours on weekdays (suggesting commuters).
This graphic values ordinal comparisons over magnitude judgments: while values can be looked up, doing so requires utilizing color, as magnitude is illegible as area, and text is used to describe the date and weather.
This is inline with our comments about the data, the types that can be appropriately described (ratio), and the tasks facilitated therein.

The purposes of this graphic might be equally well served by a number of alternatives. For instance, the \taco{}s could be simply replaced by shaded calendars, rather than being dual-encoded with area, the entire plot might be replaced with a line series, or with something specifically tuned to calendar displays\cite{van1999cluster, hartl2008visualization}.
Yet the novelty, and hence possibly the visual appeal, imparted to this graphic by the use of an unusual and somewhat perceptually difficult chart form may be diminished through the use of more pedestrian graphics.
This is emblematic of the \taco{}s general use case: as a mechanism for supporting entertainment and engagement, and not as a tool for decision making.

\section{Discussion}

We now synthesize our findings, discuss our methodology, highlight limitations, and note opportunities for future work.

\subsection{What are Table Cartograms Good For?}

\taco{}s offer an intriguing combination of affordances. We argued that they support \task{Sort} and \task{Find Extremum} tasks, as well as \task{Find Anomalies} and distribution tasks under some conditions.
We saw that they can be effective for \task{Retrieve Value} tasks, which can be obscured if the cardinality of rows and columns is too large (suggesting that tables with a small axial cardinality are preferable).
We claimed that they are best suited to tables with ordinal axes with a limited range and typically with ratio data.
While not every insight in our analysis will be surprising to those familiar with geographic cartograms or tree maps, the way in which we arrived at them (through the consideration of the \taco{}'s symmetries) facilitated a broad and self-consistent description of those conclusions, as well as a pragmatically-organized account of relevant guidelines.
Yet, as we saw, \taco{}s are not without deficits.

\parahead{Weaknesses}
\taco{}s possess a prominent collection of weaknesses, which when coupled with their unusual and inconsistent visual form, severely limits their applicability.
Their primary method of encoding data (as area) is far from the most accurate perceptual channel for encoding quantitative information
\cite{mackinlay1986automating}.
They face all of the same troubles as geographic cartograms, including that they are difficult to interpret, and possess area perception challenges \cite{nusrat2017cartogram}.
\taco{}s add further perceptual difficulties as they allow convex quadrilaterals just as readily as concave ones, such that cells that have a common area may not appear to be identical.
While their \hallucs{} allow a great degree of aesthetic control, they do not provide a consistent visual bedrock upon which to conduct analysis,
suggesting that they are ineffective for exploring data.
They are unsuitable for many tasks and datasets, such as comparisons of absolute values and datasets with wide value ranges.
In most contexts, there are an array of alternatives which may be more effective, such as tables, shaded matrices, mosaics, and tree maps.
These forms often offer preferable variations of the \taco{}'s strengths.
For instance, \taco{}s support \task{Find Anomalies}, yet tree maps facilitate this task as well \cite{munzner2014visualization} while affording the easier area comparisons.
This suggests that effective usage may involve leveraging \taco{}s' more unusual maintenance of adjacency.
However, in most contexts adjacency is not an important property to preserve at the expense of legibility.

\parahead{Potential Opportunities}
Despite these weaknesses, \taco{}s may yet be useful in an appropriate context.
Visual complexity can increase viewer engagement or enjoyment \cite{limacomplexity, harrison2015infographic}.
Hullman \etal \cite{hullman2011benefitting} argue that graphics that are visually difficult can be usefully deployed as a design element to prompt engagement.
Kosara~\cite{kosara2016presentation} suggests that presentation-only tasks can leverage aesthetic appeal for reader engagement, as with pie charts.
The atypical shapes assumed by \taco{}s may help capture readers' attention, although further study is required to evaluate this assertion.
To this end, they may be effective at capturing \emph{enjoy} tasks\cite{brehmer2013multi}, which is inline with the advocacy and education applications that geographic cartograms typically serve \cite{nusrat2016state,tobler2004thirty}.

The naturally bidirectional ordinal domain of many displays of time offers a compelling application for \taco{}s.
In \secref{sec:data-analysis} we argued that only flat tables with ordinal data should be used as inputs to \taco{}s, which are well matched displays of time, just as in calendars.
The ordered nature of calendars allows \taco{}s to highlight temporal anomalies and trends, such as in \figref{fig:cal-diff} and \secref{sec:case-study}.
This is inline with Drucker's call for ``graphical displays that emphasize the relational and co-dependent quality of temporal events'' \cite{drucker2011humanities}, as \taco{}s intermingle the structure of space with data.
However, even this algebraically-sound application should be handled with care, as month calendars have a small \halluc{} due to the ambiguity of which day of the week is first.

\parahead{Future work}
Several questions remain about \taco{}s, including the human response to them and how our recommendations perform in real contexts, each of which would benefit from a user study.
We ignored questions of interactivity (a limitation imposed by AVD), but future work might explore how animation and interactivity expand the design space---an  active area or research for geographic cartograms\cite{duncan2020task}.
Many of the geographic cartogram's ills can be addressed through ``good design choices''\cite{nusrat2016state}, such as legends and annotations.
The same may be true for \taco{}s, although further design exploration is necessary to understand the role these components could play.

\subsection{How Might We Examine Unknown Visual Forms?}

Understanding appropriate usage for novel visualizations is an intrinsically difficult, yet important task.
Previous works\cite{zuk2006theoretical} argued for theory-based analysis methods which would allow for the discount evaluation of graphics that have not become sufficiently developed or well-known to prompt user studies. Such economical methods have been shown to have great value in other contexts, such as usability testing \cite{bangor2009determining} or evaluation \cite{green1989cognitive}.

In this work we exemplified one such method of abstract evaluation based on the language of \emph{Algebraic Visualization Design}, and, in doing so, we showed that practical guidelines can be generated through application of theory.
AVD provided particular value in this pursuit, as it offered a systematic way to ask and answer questions about the visual form of interest. While not able to address every concern---most matters related to perceptual quality being too subtle for the coarse way in which we applied AVD---it was able to help us build a coherent set of usage suggestions.
This approach inverts AVD's typical usage as guide in the design process, wherein designs are invalidated for a particular task \cite{correll2018looks, crisan2021user, pu2020probabilistic, wood2018design}. Similar analyses to ours have been used to consider novel graphics\cite{correll2016surprise, wood2008spatially}, although in an ad hoc and implicit way. We formalize this approach as \emph{Algebraic Visualization Analysis} (AVA).

AVA addresses two key questions (\figref{fig:methodology-fig}).
The first---\emph{What data can be used?}---is answered  by developing a model of possible inputs (across type, structure, and size), then pruning that space via adversarial examples, whose effects were considered through the lens of AVD's \confusers{} and \hallucs{}.
Our model of possible inputs was dictated by the form of data that \taco{}s accept. Analysis of another graphic, such as a pie chart, would require a model incorporating input data, aggregation, and other relevant parameters.
We used a variety of tools in this trimming (including  Stevens' data types \cite{stevens1946theory} and JNDs), which were selected to probe the \taco{}'s particular properties. Analysis of a different graphic may necessitate a different selection of tools.
For instance, analysis of a pie chart may involve a calculation of the minimum perceptible arc area, as well as consideration of the effect of affine-versus-ratio value transformations.

The second question---\emph{What tasks can be performed?}--- is addressed by again segmenting a possibility space, this time using Amar \etals{}\cite{amar2005low} task taxonomy and then constructing relevant $\alpha$s for each task.
AVA is composed of various interlocking theories, many of which could be swapped, should the need arise. For instance, this taxonomy could be exchanged for another---although we favor this one for its brevity.
Analysis of a different chart form would involve identifying relevant $\alpha$s for each task and visually evaluating them. Future work might develop tools that given a formal specification of allowed data derive  $\alpha$s pre-matched to tasks.

Following Hu \etals \cite{hu2019viznet} triplet model of visualization a third potential question might ask \emph{What visualization is appropriate?} However, as selection of a visual form is prior to AVA, this question might be refined to \emph{How do this visual form's non-data parameters affect its usage?}
This arises in our discussion of \taco{}'s multiplicity of solutions for a given dataset.
Such issues can be answered with AVD, such as in Correll \etals \cite{correll2018looks} study of the effect that bin-widths have on histograms.
These matters are prior to AVA's key questions, as developing an understanding of what forms a chart might take on is vital to probing its usage.

The goal of answering these questions (\emph{providing chart making guidance}) is one shared by a variety of mediums,
such as guidelines\cite{diehl2020studying},  validations\cite{mcnutt2018linting, hopkins2020visualint}, prompts for introspection\cite{dork2013critical,wood2018design}, and recommendations \cite{lee_insight_2020}.
We believe that advances in any of these interwoven modalities should lead to advances in the others, and
that AVD is an opportune foundation on which  to unify them, as it is both human and machine\cite{crisan2021user, veras2019discriminability} operable.
Yet, such research directions should be followed with caution as both AVA and AVD have limitations.

\parahead{AVA Limitations}
As with any theory, AVA has a set of limitations and imprecisions.
Any investigation conducted by a single individual will likely exhibit bias. While this is problematic, the nature of our approach is somewhat self-regulating. This approach does not verify that a given visual form is good for a certain task: rather, it only invalidates particular tasks or inputs for that form.
We believe that AVA's non-existence proofs offer a valuable picture of the graphic under consideration while usefully limiting the scope of claims that can be made.
These assertions can be checked by simply comparing outputs across $\alpha$s.
AVA's iterated partitioning can inadvertently ignore errors outside of it's framing. However we do not claim this analysis to be total, instead we claim that it upper bounds performance.
Evocatively: we see a graphic as being as bad as AVA describes it to be, or worse.
In addition to the \confusers{} and \hallucs{}, AVD has two further failure modes: \textbf{\emph{Jumblers}} and \textbf{\emph{Misleaders}},
which test whether an $\alpha$ \emph{appropriately} corresponds to an $\omega$.
While they can provide useful insights, they rely on more observer-dependent judgements than the modes we considered.

Visualizations are situated in their context and their value is dependent on how they perform at tasks in those situations.
In our analysis we considered a visual form devoid of such a context
and reduced it to its most elementary components.
Our hope is that by considering the form in the abstract, that the rendered guidelines will be generally applicable. Yet, this reductive approach may cause our suggestions to be invalid in some contexts.
In future work we will compare it with other analysis methods for graphics whose properties are already well known, so as to develop a deeper understanding of what can and cannot be achieved with AVA.

We encountered phenomena in our analysis that could not be addressed through our limited set of tools.
As noted in \secref{sec:data-analysis}, while coherent guidelines can be produced using this framing, the precision of such recommendations may be limited.
Echoing Chen \etal \cite{chen2010information}, we believe that user studies are necessary for the rigorous study of visualizations (and therein construction of guidance), and that theory cannot stand alone.
The dual of this statement---that user studies must be built atop theory---does not hold, as valid experiments can be conducted without an overarching theoretical foundation.
That said, we believe that theory-based analysis has value.
Beyond its demonstrated value for discount evaluations, it can, for instance, support hypothesis generation, the result of which can be used as the basis for experiments.

\parahead{AVD Limitations}
While AVD can be an advantageous framework on which to base analyses, its structural limitations can impede some analyses.
The assertions it makes are coarse-grained, which---while being readily interpretable---can fail to provide an explanation for subtler phenomena (such as dual encodings).
It can demonstrate that tasks \emph{can be achieved}, yet it does not offer any explanation on \emph{how easily} those tasks might be under-taken.
This is analogous to how heuristic evaluation can highlight problems, but cannot suggest repairs \cite{nielsen1990heuristic}.
It is unable to validate interactive or time varying graphics (such as HOP plots\cite{kale2018hypothetical}).
It cannot reason about design components that are not purely graphical, such as embodied affect\cite{d2016feminist} (as in chartjunk\cite{bateman2010useful}) or value-sensitivity.  Schwabish and Feng describe how a line chart for a race-based COVID chart gave way to deficiency framings of those races most affected\cite{schwabish2020applying}.
While AVD might consider such an issue by generating $\alpha$s related to race, judgement of the rendered change is not limited purely to the visual spectrum as other $\alpha$s tend to be; instead it relies on the composition of biases in the viewer for validation. This may prevent the analyst from having sufficient distance to judge the sensitivity of their construction.
Constructing analysis frameworks that are value-sensitive is an important task, and should be a component of future work.
While there has been some examination of the connection between AVD's failure modes and human perception\cite{correll2018looks}, questions remain.
To accept a theory as a basis of analysis it is important to demonstrate that it has a clear connection with the real world.
AVD's explanatory power seen here and elsewhere, suggests it can validly answer questions, although experimental data is still outstanding.

\section{Acknowledgments}
We thank our reviewers for constructive feedback.
We also thank Will Brackenbury, Ravi Chugh, Brian Hempel, Gordon Kindlmann, as well as the UChicago Vis Reading group for their insightful comments, thoughtful discussion, and splendid support.

\bibliographystyle{eg-alpha-doi}
\bibliography{table-cartograms}

\newcommand{\etalchar}[1]{$^{#1}$}
\begin{thebibliography}{\uppercase{VDBBC{\etalchar{*}}18}}

\bibitem[AES05]{amar2005low}
\textsc{Amar R., Eagan J., Stasko J.}:
\newblock {Low-Level Components of Analytic Activity in Information
  Visualization}.
\newblock In \emph{IEEE Symposium on Information Visualization} (2005), IEEE,
  pp.~111--117.
\newblock \href {https://doi.org/10.1109/INFVIS.2005.1532136}
  {\path{doi:10.1109/INFVIS.2005.1532136}}.

\bibitem[AKV15]{alam2015quantitative}
\textsc{Alam M.~J., Kobourov S.~G., Veeramoni S.}:
\newblock {Quantitative Measures for Cartogram Generation Techniques}.
\newblock In \emph{Computer Graphics Forum} (2015), vol.~34, Wiley Online
  Library, pp.~351--360.
\newblock \href {https://doi.org/10.1111/cgf.12647}
  {\path{doi:10.1111/cgf.12647}}.

\bibitem[AL20]{adar2020communicative}
\textsc{Adar E., Lee E.}:
\newblock {Communicative Visualizations as a Learning Problem}.
\newblock \emph{IEEE Transactions on Visualization and Computer Graphics}
  (2020).

\bibitem[Ans73]{anscomb}
\textsc{Anscombe F.}:
\newblock Graphs in statistical analysis.
\newblock \emph{The American Statistician 27}, 1 (1973), 17--21.

\bibitem[BBD20]{bruggemann2020Fold}
\textsc{Br{\"u}ggemann V., Bludau M.-J., D{\"o}rk M.}:
\newblock {The Fold: Rethinking Interactivity in Data Visualization}.
\newblock \emph{Digital Humanities Quarterly 14}, 3 (2020).

\bibitem[BBK{\etalchar{*}}18]{behrisch2018quality}
\textsc{Behrisch M., Blumenschein M., Kim N.~W., Shao L., El-Assady M., Fuchs
  J., Seebacher D., Diehl A., Brandes U., Pfister H., et~al.}:
\newblock {Quality Metrics for Information Visualization}.
\newblock In \emph{Computer Graphics Forum} (2018), vol.~37, Wiley Online
  Library, pp.~625--662.
\newblock \href {https://doi.org/10.1111/cgf.13446}
  {\path{doi:10.1111/cgf.13446}}.

\bibitem[BCF20]{bertini2020shouldn}
\textsc{Bertini E., Correll M., Franconeri S.}:
\newblock {Why Shouldn't All Charts Be Scatter Plots? Beyond Precision-Driven
  Visualizations}.
\newblock In \emph{IEEE Visualization Conference (VIS)} (2020), IEEE, pp.~1--5.
\newblock \href {https://doi.org/10.1109/VIS47514.2020.00048}
  {\path{doi:10.1109/VIS47514.2020.00048}}.

\bibitem[BKM09]{bangor2009determining}
\textsc{Bangor A., Kortum P., Miller J.}:
\newblock {Determining What Individual SUS Scores Mean: Adding an Adjective
  Rating Scale}.
\newblock \emph{Journal of usability studies 4}, 3 (2009), 114--123.

\bibitem[BKM13]{bakke2013automatic}
\textsc{Bakke E., Karger D.~R., Miller R.~C.}:
\newblock {Automatic Layout of Structured Hierarchical Reports}.
\newblock \emph{{IEEE Transactions on Visualization and Computer Graphics} 19},
  12 (2013), 2586--2595.
\newblock \href {https://doi.org/10.1109/TVCG.2013.137}
  {\path{doi:10.1109/TVCG.2013.137}}.

\bibitem[BM13]{brehmer2013multi}
\textsc{Brehmer M., Munzner T.}:
\newblock {A Multi-Level Typology of Abstract Visualization Tasks}.
\newblock \emph{{IEEE Transactions on Visualization and Computer Graphics} 19},
  12 (2013), 2376--2385.
\newblock \href {https://doi.org/10.1109/TVCG.2013.124}
  {\path{doi:10.1109/TVCG.2013.124}}.

\bibitem[BMG{\etalchar{*}}10]{bateman2010useful}
\textsc{Bateman S., Mandryk R.~L., Gutwin C., Genest A., McDine D., Brooks C.}:
\newblock {Useful Junk? The Effects of Visual Embellishment on Comprehension
  and Memorability of Charts}.
\newblock In \emph{Proceedings of the SIGCHI Conference on Human Factors in
  Computing Systems} (2010), pp.~2573--2582.
\newblock \href {https://doi.org/10.1145/1753326.1753716}
  {\path{doi:10.1145/1753326.1753716}}.

\bibitem[Bra20]{brath2020literal}
\textsc{Brath R.}:
\newblock {Literal Encoding: Text is a first-class data encoding}.
\newblock \emph{Visualization for the Digital Humanities (VIS4DH)} (2020).

\bibitem[BZJ{\etalchar{*}}20]{bares2020using}
\textsc{Bares A., Zeller S., Jackson C.~D., Keefe D.~F., Samsel F.}:
\newblock Using close reading as a method for evaluating visualizations.
\newblock In \emph{{IEEE} Workshop on Evaluation and Beyond - Methodological
  Approaches to Visualization, 2020} (2020), {IEEE}, pp.~29--37.
\newblock \href {https://doi.org/10.1109/BELIV51497.2020.00011}
  {\path{doi:10.1109/BELIV51497.2020.00011}}.

\bibitem[CC21]{crisan2021user}
\textsc{Crisan A., Correll M.}:
\newblock {User Ex Machina: Simulation as a Design Probe in Human-in-the-Loop
  Text Analytics}.
\newblock In \emph{Proceedings of the SIGCHI Conference on Human Factors in
  Computing Systems} (2021), pp.~1--16.
\newblock To Appear.

\bibitem[CG15]{chen2015may}
\textsc{Chen M., Golan A.}:
\newblock {What May Visualization Processes Optimize?}
\newblock \emph{{IEEE Transactions on Visualization and Computer Graphics} 22},
  12 (2015), 2619--2632.
\newblock \href {https://doi.org/10.1109/TVCG.2015.2513410}
  {\path{doi:10.1109/TVCG.2015.2513410}}.

\bibitem[CH16]{correll2016surprise}
\textsc{Correll M., Heer J.}:
\newblock {Surprise! Bayesian Weighting for De-Biasing Thematic Maps}.
\newblock \emph{{IEEE Transactions on Visualization and Computer Graphics} 23},
  1 (2016), 651--660.
\newblock \href {https://doi.org/10.1109/TVCG.2016.2598618}
  {\path{doi:10.1109/TVCG.2016.2598618}}.

\bibitem[CH17]{blackhat}
\textsc{Correll M., Heer J.}:
\newblock {Black Hat Visualization}.
\newblock In \emph{{DECISIVe : Workshop on Dealing with Cognitive Biases in
  Visualisations}} (2017).

\bibitem[{Chi}a]{chicago-red-light}
\textsc{{Chicago Data Portal}}:
\newblock {Speed Camera Violations}.
\newblock
  \webLink{https://data.cityofchicago.org/Transportation/Speed-Camera-Violations/hhkd-xvj4}.
\newblock Accessed 09/23/2018.

\bibitem[{Chi}b]{cps-schedule}
\textsc{{Chicago Public Schools}}:
\newblock {{School Year Calendar}}.
\newblock \webLink{www.cps.edu}.
\newblock Accessed 10/04/2018.

\bibitem[CJ10]{chen2010information}
\textsc{Chen M., Jaenicke H.}:
\newblock {An Information-theoretic Framework for Visualization}.
\newblock \emph{{IEEE Transactions on Visualization and Computer Graphics} 16},
  6 (2010), 1206--1215.
\newblock \href {https://doi.org/10.1109/TVCG.2010.132}
  {\path{doi:10.1109/TVCG.2010.132}}.

\bibitem[CLKS18]{correll2018looks}
\textsc{Correll M., Li M., Kindlmann G., Scheidegger C.}:
\newblock {Looks Good To Me: Visualizations As Sanity Checks}.
\newblock \emph{{IEEE Transactions on Visualization and Computer Graphics} 25},
  1 (2018), 830--839.
\newblock \href {https://doi.org/10.1109/TVCG.2018.2864907}
  {\path{doi:10.1109/TVCG.2018.2864907}}.

\bibitem[CM84]{Cleveland-GraphPercept-1984}
\textsc{Cleveland W.~S., McGill R.}:
\newblock {Graphical Perception: Theory, Experimentation, and Application to
  the Development of Graphical Methods}.
\newblock \emph{{Journal of the American Statistical Association}} (1984).

\bibitem[CM86]{cleveland1986experiment}
\textsc{Cleveland W.~S., McGill R.}:
\newblock {An Experiment in Graphical Perception}.
\newblock \emph{International Journal of Man-Machine Studies 25}, 5 (1986),
  491--500.

\bibitem[CM07]{cawthon2007effect}
\textsc{Cawthon N., Moere A.~V.}:
\newblock {The Effect of Aesthetic on the Usability of Data Visualization}.
\newblock In \emph{11th International Conference Information Visualization}
  (2007), IEEE, pp.~637--648.
\newblock \href {https://doi.org/10.1109/IV.2007.147}
  {\path{doi:10.1109/IV.2007.147}}.

\bibitem[DFCC13]{dork2013critical}
\textsc{D{\"o}rk M., Feng P., Collins C., Carpendale S.}:
\newblock {Critical InfoVis: exploring the politics of visualization}.
\newblock In \emph{Extended Abstracts on Human Factors in Computing Systems}.
  ACM, 2013, pp.~2189--2198.
\newblock \href {https://doi.org/10.1145/2468356.2468739}
  {\path{doi:10.1145/2468356.2468739}}.

\bibitem[DK16]{d2016feminist}
\textsc{D’Ignazio C., Klein L.~F.}:
\newblock {Feminist Data Visualization}.
\newblock In \emph{Workshop on Visualization for the Digital Humanities
  (VIS4DH), Baltimore. IEEE} (2016).

\bibitem[DKA{\etalchar{*}}20]{diehl2020studying}
\textsc{Diehl A., Kraus M., Abdul{-}Rahman A., El{-}Assady M., Bach B., Laramee
  R.~S., Keim D.~A., Chen M.}:
\newblock {Studying Visualization Guidelines According to Grounded Theory}.
\newblock URL: \url{https://arxiv.org/abs/2010.09040}, \href
  {http://arxiv.org/abs/2010.09040} {\path{arXiv:2010.09040}}.

\bibitem[Dru11]{drucker2011humanities}
\textsc{Drucker J.}:
\newblock {Humanities Approaches to Graphical Display}.
\newblock \emph{Digital Humanities Quarterly 5}, 1 (2011), 1--21.

\bibitem[DSK{\etalchar{*}}14]{demiralp2014visual}
\textsc{Demiralp {\c{C}}., Scheidegger C.~E., Kindlmann G.~L., Laidlaw D.~H.,
  Heer J.}:
\newblock {Visual Embedding: {A} Model for Visualization}.
\newblock \emph{{IEEE Computer Graphics and Applications} 34}, 1 (2014),
  10--15.
\newblock \href {https://doi.org/10.1109/MCG.2014.18}
  {\path{doi:10.1109/MCG.2014.18}}.

\bibitem[DTPG21]{duncan2020task}
\textsc{Duncan I.~K., Tingsheng S., Perrault S.~T., Gastner M.~T.}:
\newblock {Task-Based Effectiveness of Interactive Contiguous Area Cartograms}.
\newblock \emph{{IEEE} Transactions on Visualization and Computer Graphics 27},
  3 (2021), 2136--2152.
\newblock \href {https://doi.org/10.1109/TVCG.2020.3041745}
  {\path{doi:10.1109/TVCG.2020.3041745}}.

\bibitem[EFK{\etalchar{*}}13]{evans2013table}
\textsc{Evans W., Felsner S., Kaufmann M., Kobourov S.~G., Mondal D., Nishat
  R.~I., Verbeek K.}:
\newblock {Table Cartograms}.
\newblock In \emph{European Symposium on Algorithms} (2013), Springer,
  pp.~421--432.
\newblock \href {https://doi.org/10.1007/978-3-642-40450-4\_36}
  {\path{doi:10.1007/978-3-642-40450-4\_36}}.

\bibitem[EvKSS15]{eppstein2015improved}
\textsc{Eppstein D., van Kreveld M., Speckmann B., Staals F.}:
\newblock {Improved Grid Map Layout by Point Set Matching}.
\newblock \emph{International Journal of Computational Geometry \& Applications
  25}, 02 (2015), 101--122.

\bibitem[FAAa]{senateWebsite}
\textsc{FAA}:
\newblock {United States Senate}.
\newblock \webLink{https://www.senate.gov}.

\bibitem[FAAb]{faa-strike}
\textsc{FAA}:
\newblock {{Wildlife Strike Database}}.
\newblock \webLink{https://wildlife.faa.gov}.

\bibitem[Fri94]{friendly1994mosaic}
\textsc{Friendly M.}:
\newblock {Mosaic Displays for Multi-Way Contingency Tables}.
\newblock \emph{Journal of the American Statistical Association 89}, 425
  (1994), 190--200.

\bibitem[Gre89]{green1989cognitive}
\textsc{Green T.~R.}:
\newblock {Cognitive Dimensions of Notations}.
\newblock \emph{People and Computers V} (1989), 443--460.

\bibitem[Har08]{hartl2008visualization}
\textsc{Hartl P.~R.}:
\newblock \emph{Visualization of Calendar Data}.
\newblock PhD thesis, Vienna University of Technology, 2008.

\bibitem[HAS11]{hullman2011benefitting}
\textsc{Hullman J., Adar E., Shah P.}:
\newblock {Benefitting Infovis with Visual Difficulties}.
\newblock \emph{{IEEE Transactions on Visualization and Computer Graphics}}
  (2011).
\newblock \href {https://doi.org/10.1109/TVCG.2011.175}
  {\path{doi:10.1109/TVCG.2011.175}}.

\bibitem[Has21]{Hasan2021}
\textsc{Hasan M.~R.}:
\newblock tcarto\_applications.
\newblock \url{https://github.com/rakib045/tcarto_applications}, 2021.
\newblock Accessed 2/21/2021.

\bibitem[HCS20]{hopkins2020visualint}
\textsc{Hopkins A.~K., Correll M., Satyanarayan A.}:
\newblock {VisuaLint: Sketchy In Situ Annotations of Chart Construction
  Errors}.
\newblock In \emph{Computer Graphics Forum} (2020), vol.~39, Wiley Online
  Library, pp.~219--228.
\newblock \href {https://doi.org/10.1111/cgf.13975}
  {\path{doi:10.1111/cgf.13975}}.

\bibitem[HGH{\etalchar{*}}19]{hu2019viznet}
\textsc{Hu K., Gaikwad S., Hulsebos M., Bakker M.~A., Zgraggen E., Hidalgo C.,
  Kraska T., Li G., Satyanarayan A., Demiralp {\c{C}}.}:
\newblock {{VizNet: Towards {A} Large-Scale Visualization Learning and
  Benchmarking Repository}}.
\newblock In \emph{Proceedings of the SIGCHI Conference on Human Factors in
  Computing Systems} (2019), pp.~1--12.
\newblock \href {https://doi.org/10.1145/3290605.3300892}
  {\path{doi:10.1145/3290605.3300892}}.

\bibitem[HRC15]{harrison2015infographic}
\textsc{Harrison L., Reinecke K., Chang R.}:
\newblock {Infographic Aesthetics: Designing for the First Impression}.
\newblock In \emph{Proceedings of the SIGCHI Conference on Human Factors in
  Computing Systems} (2015).
\newblock \href {https://doi.org/10.1145/2702123.2702545}
  {\path{doi:10.1145/2702123.2702545}}.

\bibitem[Hur06]{hurst2006towards}
\textsc{Hurst M.}:
\newblock {Towards a Theory of Tables}.
\newblock \emph{International Journal of Document Analysis and Recognition 8},
  2-3 (2006), 123--131.
\newblock \href {https://doi.org/10.1007/s10032-006-0016-y}
  {\path{doi:10.1007/s10032-006-0016-y}}.

\bibitem[IL20]{inoue2020optimization}
\textsc{Inoue R., Li M.}:
\newblock {Optimization-Based Construction of Quadrilateral Table Cartograms}.
\newblock \emph{ISPRS International Journal of Geo-Information 9}, 1 (2020),
  43.
\newblock \href {https://doi.org/10.3390/ijgi9010043}
  {\path{doi:10.3390/ijgi9010043}}.

\bibitem[KNKH18]{kale2018hypothetical}
\textsc{Kale A., Nguyen F., Kay M., Hullman J.}:
\newblock {Hypothetical Outcome Plots Help Untrained Observers Judge Trends in
  Ambiguous Data}.
\newblock \emph{{IEEE Transactions on Visualization and Computer Graphics} 25},
  1 (2018), 892--902.
\newblock \href {https://doi.org/10.1109/TVCG.2018.2864909}
  {\path{doi:10.1109/TVCG.2018.2864909}}.

\bibitem[Kos10]{kosara-pie-charts}
\textsc{Kosara R.}:
\newblock {{Understanding Pie Charts}}.
\newblock \webLink{https://eagereyes.org/techniques/pie-charts}, 2010.
\newblock Accessed 11/2/2020.

\bibitem[Kos16]{kosara2016presentation}
\textsc{Kosara R.}:
\newblock {Presentation-Oriented Visualization Techniques}.
\newblock \emph{{IEEE Computer Graphics and Applications} 36}, 1 (2016),
  80--85.
\newblock \href {https://doi.org/10.1109/MCG.2016.2}
  {\path{doi:10.1109/MCG.2016.2}}.

\bibitem[KRK01]{krider2001pizzas}
\textsc{Krider R.~E., Raghubir P., Krishna A.}:
\newblock {Pizzas: $\pi$ or Square? Psychophysical Biases in Area Comparisons}.
\newblock \emph{Marketing Science 20}, 4 (2001), 405--425.
\newblock \href {https://doi.org/10.1287/mksc.20.4.405.9756}
  {\path{doi:10.1287/mksc.20.4.405.9756}}.

\bibitem[KS14]{kindlmann2014algebraic}
\textsc{Kindlmann G., Scheidegger C.}:
\newblock {An Algebraic Process for Visualization Design}.
\newblock \emph{{IEEE Transactions on Visualization and Computer Graphics} 20},
  12 (2014), 2181--2190.
\newblock \href {https://doi.org/10.1109/TVCG.2014.2346325}
  {\path{doi:10.1109/TVCG.2014.2346325}}.

\bibitem[KS16]{KindlmannAlgebraicVisPedagogyPDV2016}
\textsc{Kindlmann G., Scheidegger C.}:
\newblock {Algebraic Visualization Design for Pedagogy}.
\newblock IEEE VIS Workshop on Pedagogy of Data Visualization, Oct. 2016.

\bibitem[Lee20]{lee_insight_2020}
\textsc{Lee D. J.-L.}:
\newblock Insight {Machines}: {The} {Past}, {Present}, and {Future} of
  {Visualization} {Recommendation}.
\newblock \webLink{https://link.medium.com/fM5VhrSL5db}, Feb. 2020.
\newblock Multiple Views: Visualization Research Explained. Accessed 2/21/2021.

\bibitem[LI19]{li2019table}
\textsc{Li M., Inoue R.}:
\newblock {Table Cartogram Generation as an Optimization Problem}.
\newblock \emph{Abstracts of the International Cartographic Association 1}
  (2019), 1--2.

\bibitem[Lim11]{limacomplexity}
\textsc{Lima M.}:
\newblock \emph{{Visual Complexity: Mapping Patterns of Information}}.
\newblock Princeton Architectural Press, 2011.

\bibitem[Mac86]{mackinlay1986automating}
\textsc{Mackinlay J.}:
\newblock {Automating the Design of Graphical Presentations of Relational
  Information}.
\newblock \emph{ACM Transactions On Graphics} (1986).
\newblock \href {https://doi.org/10.1145/22949.22950}
  {\path{doi:10.1145/22949.22950}}.

\bibitem[MCC20]{mcnutt2020divining}
\textsc{McNutt A., Crisan A., Correll M.}:
\newblock Divining insights: Visual analytics through cartomancy.
\newblock In \emph{Extended Abstracts of the CHI Conference on Human Factors in
  Computing Systems} (2020), pp.~1--16.
\newblock \href {https://doi.org/10.1145/3334480.3381814}
  {\path{doi:10.1145/3334480.3381814}}.

\bibitem[McN19]{mcnutt2019Cycles}
\textsc{McNutt A.}:
\newblock {Cycles Rain Seasons In Size}.
\newblock
  \webLink{https://www.informationisbeautifulawards.com/showcase/4499-cycles-rain-seasons-in-size},
  2019.
\newblock Accessed 11/2/2020.

\bibitem[MD73]{martinez1973ranking}
\textsc{Martinez N., Dawson W.~E.}:
\newblock {Ranking of Apparent Area for Different Shapes of Equal Area}.
\newblock \emph{{Perceptual and Motor Skills}} (1973).
\newblock \href {https://doi.org/10.1177/003151257303700319}
  {\path{doi:10.1177/003151257303700319}}.

\bibitem[{Met}]{ohare-weather}
\textsc{{Meteoblue}}:
\newblock {{O'hare International Airport Weather}}.
\newblock
  \webLink{www.meteoblue.com/en/weather/archive/export/chicago-o\%27hare-international-airport_united-states-of-america_4887479}.
\newblock Accessed 09/23/2018.

\bibitem[MK18]{mcnutt2018linting}
\textsc{McNutt A., Kindlmann G.}:
\newblock {Linting for Visualization: Towards a Practical Automated
  Visualization Guidance System}.
\newblock In \emph{VisGuides: 2nd Workshop on the Creation, Curation, Critique
  and Conditioning of Principles and Guidelines in Visualization} (2018).

\bibitem[MK20]{mcnutt2020Minimally}
\textsc{McNutt A., Kindlmann G.}:
\newblock {A Minimally Constrained Optimization Algorithm for Table
  Cartograms}.
\newblock \emph{IEEEVIS InfoVis Posters} (2020).

\bibitem[MKC20]{mcnutt2020surfacing}
\textsc{McNutt A.~M., Kindlmann G.~L., Correll M.}:
\newblock {Surfacing Visualization Mirages}.
\newblock In \emph{Proceedings of the SIGCHI Conference on Human Factors in
  Computing Systems} (2020), {ACM}, pp.~1--16.
\newblock \href {https://doi.org/10.1145/3313831.3376420}
  {\path{doi:10.1145/3313831.3376420}}.

\bibitem[Mun14]{munzner2014visualization}
\textsc{Munzner T.}:
\newblock \emph{{Visualization Analysis and Design}}.
\newblock AK Peters/CRC Press, 2014.

\bibitem[{Nat}]{zion-stats}
\textsc{{National Park Service}}:
\newblock {Zion Park Visitation Statistics}.
\newblock
  \webLink{https://www.nps.gov/zion/learn/management/park-visitation-statistics.htm}.
\newblock Accessed 10/04/2018.

\bibitem[NK15]{nusratCartogramTask}
\textsc{Nusrat S., Kobourov S.}:
\newblock {Task Taxonomy for Cartograms}.
\newblock In \emph{Proceedings of the Eurographics / IEEE VGTC Conference on
  Visualization: Short Papers} (2015).
\newblock \href {https://doi.org/10.2312/eurovisshort.20151126}
  {\path{doi:10.2312/eurovisshort.20151126}}.

\bibitem[NK16]{nusrat2016state}
\textsc{Nusrat S., Kobourov S.}:
\newblock {The State of the Art in Cartograms}.
\newblock In \emph{Computer Graphics Forum} (2016), vol.~35, pp.~619--642.
\newblock \href {https://doi.org/10.1111/cgf.12932}
  {\path{doi:10.1111/cgf.12932}}.

\bibitem[NM90]{nielsen1990heuristic}
\textsc{Nielsen J., Molich R.}:
\newblock {Heuristic Evaluation of User Interfaces}.
\newblock In \emph{Proceedings of the SIGCHI Conference on Human Factors in
  Computing Systems} (1990), {ACM}, pp.~249--256.
\newblock \href {https://doi.org/10.1145/97243.97281}
  {\path{doi:10.1145/97243.97281}}.

\bibitem[Nus17]{nusrat2017cartogram}
\textsc{Nusrat S.}:
\newblock \emph{Cartogram Visualization: Methods, Applications, and
  Effectiveness}.
\newblock PhD thesis, The University of Arizona, 2017.

\bibitem[pie]{pierebean_oc_2020}
\textsc{pierebean}:
\newblock [{OC}] >19k chinese characters sorted by pronunciation.
\newblock
  \webLink{www.reddit.com/r/dataisbeautiful/comments/jm042v/oc_19k_chinese_characters_sorted_by_pronunciation/}.
\newblock Accessed 11/2/2020.

\bibitem[PK20]{pu2020probabilistic}
\textsc{Pu X., Kay M.}:
\newblock {A Probabilistic Grammar of Graphics}.
\newblock In \emph{Proceedings of the SIGCHI Conference on Human Factors in
  Computing Systems} (2020), pp.~1--13.
\newblock \href {https://doi.org/10.1145/3313831.3376466}
  {\path{doi:10.1145/3313831.3376466}}.

\bibitem[{Red}13]{reddit:dnd-alignment}
\textsc{{Reddit Users}}:
\newblock {Alignment Survey - Results}.
\newblock
  \webLink{https://www.reddit.com/r/DnD/comments/1ejnft/alignment_survey_results/},
  2013.
\newblock Accessed 11/2/2020.

\bibitem[Ros18]{dual-y-axis}
\textsc{Rost L.~C.}:
\newblock {{Why not to use two axes, and what to use instead}}.
\newblock \webLink{https://blog.datawrapper.de/dualaxis/}, 2018.
\newblock Accessed 11/2/20.

\bibitem[RTAA20]{raidvee2020perception}
\textsc{Raidvee A., Toom M., Averin K., Allik J.}:
\newblock Perception of means, sums, and areas.
\newblock \emph{{Attention, Perception, \& Psychophysics}} (2020), 1--12.
\newblock \href {https://doi.org/10.3758/s13414-019-01938-7}
  {\path{doi:10.3758/s13414-019-01938-7}}.

\bibitem[RWC19]{ritchie2019lie}
\textsc{Ritchie J., Wigdor D., Chevalier F.}:
\newblock {A Lie Reveals the Truth: Quasimodes for Task-Aligned Data
  Presentation}.
\newblock In \emph{Proceedings of the SIGCHI Conference on Human Factors in
  Computing Systems} (2019), pp.~1--13.
\newblock \href {https://doi.org/10.1145/3290605.3300423}
  {\path{doi:10.1145/3290605.3300423}}.

\bibitem[SDW09]{slingsby2009configuring}
\textsc{Slingsby A., Dykes J., Wood J.}:
\newblock {Configuring Hierarchical Layouts to Address Research Questions}.
\newblock \emph{{IEEE Transactions on Visualization and Computer Graphics} 15},
  6 (2009), 977--984.
\newblock \href {https://doi.org/10.1109/TVCG.2009.128}
  {\path{doi:10.1109/TVCG.2009.128}}.

\bibitem[SF20]{schwabish2020applying}
\textsc{Schwabish J., Feng A.}:
\newblock {Applying Racial Equity Awareness in Data Visualization}.
\newblock \emph{{Visualization for Communication}} (2020).

\bibitem[Shn96]{shneiderman1996eyes}
\textsc{Shneiderman B.}:
\newblock {The Eyes Have It: A Task by Data Type Taxonomy for Information
  Visualizations}.
\newblock In \emph{Visual Languages, 1996. Proceedings., IEEE Symposium on}
  (1996), IEEE, pp.~336--343.
\newblock \href {https://doi.org/10.1109/VL.1996.545307}
  {\path{doi:10.1109/VL.1996.545307}}.

\bibitem[SLD20]{scheibel2020survey}
\textsc{Scheibel W., Limberger D., D{\"{o}}llner J.}:
\newblock Survey of treemap layout algorithms.
\newblock In \emph{Proceedings of the 13th International Symposium on Visual
  Information Communication and Interaction} (2020), {ACM}, pp.~1:1--1:9.
\newblock \href {https://doi.org/10.1145/3430036.3430041}
  {\path{doi:10.1145/3430036.3430041}}.

\bibitem[Ste46]{stevens1946theory}
\textsc{Stevens S.~S.}:
\newblock {On the Theory of Scales of Measurement}.
\newblock \emph{Science} (1946).

\bibitem[Tob04]{tobler2004thirty}
\textsc{Tobler W.}:
\newblock {Thirty Five Years of Computer Cartograms}.
\newblock \emph{Annals of the Association of American Geographers 94}, 1
  (2004), 58--73.
\newblock \href {https://doi.org/10.1111/j.1467-8306.2004.09401004.x}
  {\path{doi:10.1111/j.1467-8306.2004.09401004.x}}.

\bibitem[{Uni}16]{census-state-to-state}
\textsc{{United States Census Bureau}}:
\newblock {State-to-State Migration Flows}.
\newblock
  \webLink{https://www.census.gov/data/tables/time-series/demo/geographic-mobility/state-to-state-migration.html/},
  2016.
\newblock Accessed 1/16/2019.

\bibitem[{Uni}19]{census-pops}
\textsc{{United States Census Bureau}}:
\newblock Decennial census of population and housing.
\newblock
  \webLink{https://www.census.gov/programs-surveys/decennial-census/data.html},
  2019.
\newblock Accessed 3/15/2019.

\bibitem[VC19]{veras2019discriminability}
\textsc{Veras R., Collins C.}:
\newblock {Discriminability Tests for Visualization Effectiveness and
  Scalability}.
\newblock \emph{{IEEE Transactions on Visualization and Computer Graphics} 26},
  1 (2019), 749--758.
\newblock \href {https://doi.org/10.1109/TVCG.2019.2934432}
  {\path{doi:10.1109/TVCG.2019.2934432}}.

\bibitem[VDBBC{\etalchar{*}}18]{van2018philosophical}
\textsc{Van Den~Berg H., Betti A., Castermans T., Koopman R., Speckmann B.,
  Verbeek K., Van~der Werf T., Wang S., Westenberg M.~A., et~al.}:
\newblock {A Philosophical Perspective on Visualization for Digital
  Humanities}.
\newblock In \emph{Visualization for the Digital Humanities (VIS4DH)} (2018).

\bibitem[VFR12]{vickers2012understanding}
\textsc{Vickers P., Faith J., Rossiter N.}:
\newblock {Understanding Visualization: {A} Formal Approach Using Category
  Theory and Semiotics}.
\newblock \emph{{IEEE Transactions on Visualization and Computer Graphics} 19},
  6 (2012), 1048--1061.
\newblock \href {https://doi.org/10.1109/TVCG.2012.294}
  {\path{doi:10.1109/TVCG.2012.294}}.

\bibitem[VWVS99]{van1999cluster}
\textsc{Van~Wijk J.~J., Van~Selow E.~R.}:
\newblock {Cluster and Calendar Based Visualization of Time Series Data}.
\newblock In \emph{IEEE Symposium on Information Visualization} (1999), IEEE,
  pp.~4--9.
\newblock \href {https://doi.org/10.1109/INFVIS.1999.801851}
  {\path{doi:10.1109/INFVIS.1999.801851}}.

\bibitem[WAB{\etalchar{*}}19]{tidyverse}
\textsc{Wickham H., Averick M., Bryan J., Chang W., McGowan L.~D., François
  R., Grolemund G., Hayes A., Henry L., Hester J., Kuhn M., Pedersen T.~L.,
  Miller E., Bache S.~M., Müller K., Ooms J., Robinson D., Seidel D.~P., Spinu
  V., Takahashi K., Vaughan D., Wilke C., Woo K., Yutani H.}:
\newblock Welcome to the {tidyverse}.
\newblock \emph{Journal of Open Source Software 4}, 43 (2019), 1686.
\newblock \href {https://doi.org/10.21105/joss.01686}
  {\path{doi:10.21105/joss.01686}}.

\bibitem[WAM{\etalchar{*}}18]{wall2018heuristic}
\textsc{Wall E., Agnihotri M., Matzen L., Divis K., Haass M., Endert A., Stasko
  J.}:
\newblock A heuristic approach to value-driven evaluation of visualizations.
\newblock \emph{{IEEE Transactions on Visualization and Computer Graphics} 25},
  1 (2018), 491--500.
\newblock \href {https://doi.org/10.1109/TVCG.2018.2865146}
  {\path{doi:10.1109/TVCG.2018.2865146}}.

\bibitem[WCHB10]{wickham2010graphical}
\textsc{Wickham H., Cook D., Hofmann H., Buja A.}:
\newblock {Graphical Inference for Infovis}.
\newblock \emph{{IEEE Transactions on Visualization and Computer Graphics} 16},
  6 (2010), 973--979.
\newblock \href {https://doi.org/10.1109/TVCG.2010.161}
  {\path{doi:10.1109/TVCG.2010.161}}.

\bibitem[WD08]{wood2008spatially}
\textsc{Wood J., Dykes J.}:
\newblock {Spatially Ordered Treemaps}.
\newblock \emph{{IEEE Transactions on Visualization and Computer Graphics} 14},
  6 (2008).
\newblock \href {https://doi.org/10.1109/TVCG.2008.165}
  {\path{doi:10.1109/TVCG.2008.165}}.

\bibitem[Wic13]{wickham2013graphical}
\textsc{Wickham H.}:
\newblock {Graphical Criticism: Some Historical Notes}.
\newblock \emph{Journal of Computational and Graphical Statistics 22}, 1
  (2013), 38--44.
\newblock \href {https://doi.org/10.1080/10618600.2012.761140}
  {\path{doi:10.1080/10618600.2012.761140}}.

\bibitem[WKD18]{wood2018design}
\textsc{Wood J., Kachkaev A., Dykes J.}:
\newblock {Design Exposition with Literate Visualization}.
\newblock \emph{{IEEE Transactions on Visualization and Computer Graphics} 25},
  1 (2018), 759--768.
\newblock \href {https://doi.org/10.1109/TVCG.2018.2864836}
  {\path{doi:10.1109/TVCG.2018.2864836}}.

\bibitem[ZC06]{zuk2006theoretical}
\textsc{Zuk T., Carpendale S.}:
\newblock {Theoretical Analysis of Uncertainty Visualizations}.
\newblock In \emph{Visualization and Data Analysis} (2006), vol.~6060,
  International Society for Optics and Photonics, p.~606007.
\newblock \href {https://doi.org/10.1117/12.643631}
  {\path{doi:10.1117/12.643631}}.

\bibitem[ZSN{\etalchar{*}}06]{zuk2006heuristics}
\textsc{Zuk T., Schlesier L., Neumann P., Hancock M.~S., Carpendale S.}:
\newblock Heuristics for information visualization evaluation.
\newblock In \emph{Proceedings of the Workshop on Beyond Time and Errors: Novel
  Evaluation Methods for Visualization, {BELIV}} (2006), {ACM} Press, pp.~1--6.
\newblock \href {https://doi.org/10.1145/1168149.1168162}
  {\path{doi:10.1145/1168149.1168162}}.

\end{thebibliography}

\clearpage
\pagebreak
\newpage
\appendix

\section{Supplemental Material}

\begin{figure}[t] 
      \centering
      \includegraphics[width=\columnwidth]{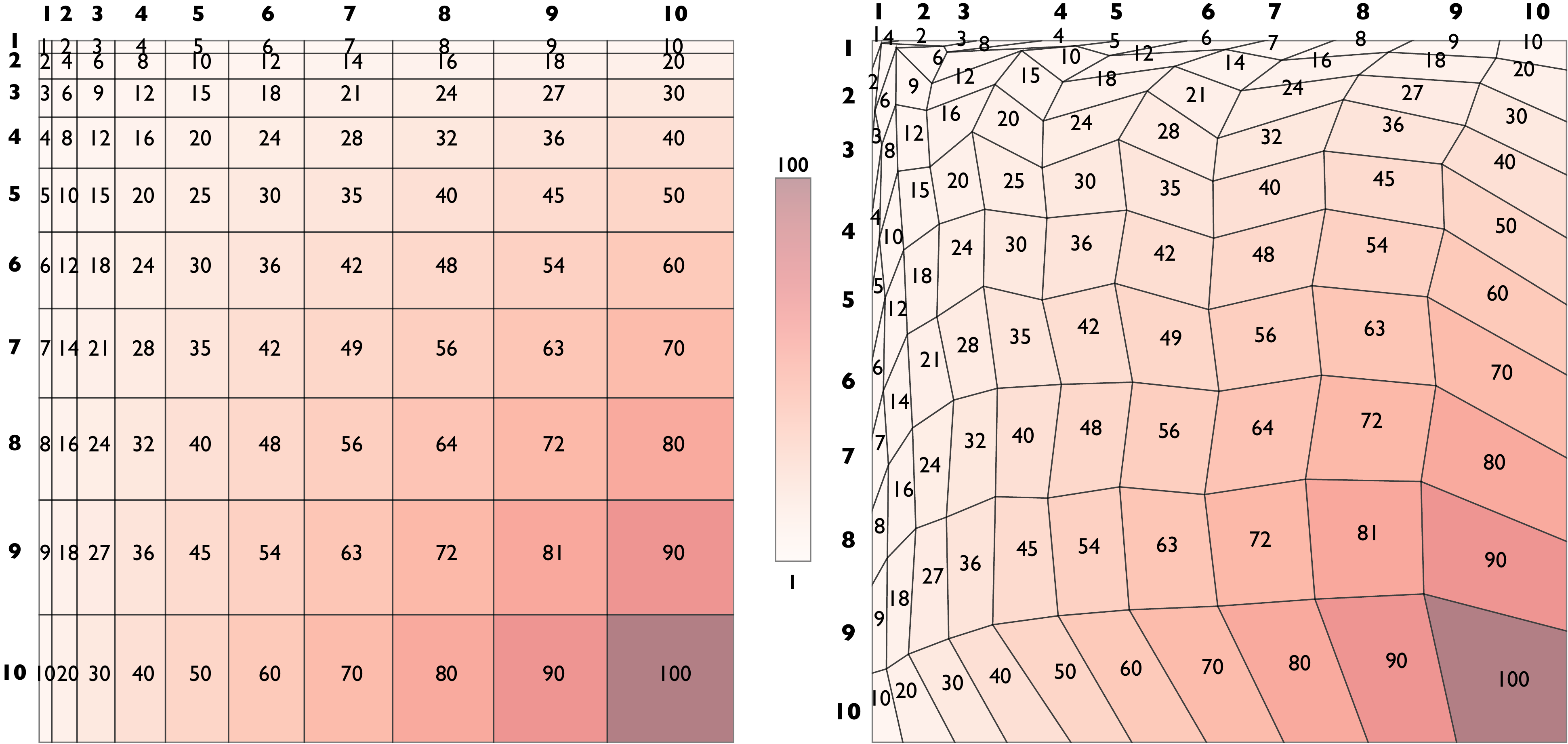}
      \caption{
            The multiplicity of accurate layouts for a single input table yields a prominent table cartogram \halluc{} as in this pair of multiplication tables.
      }
      \label{fig:multiplication}
      \spaceit
\end{figure}

\begin{figure}[t] 
    \centering
    \includegraphics[width=\columnwidth]{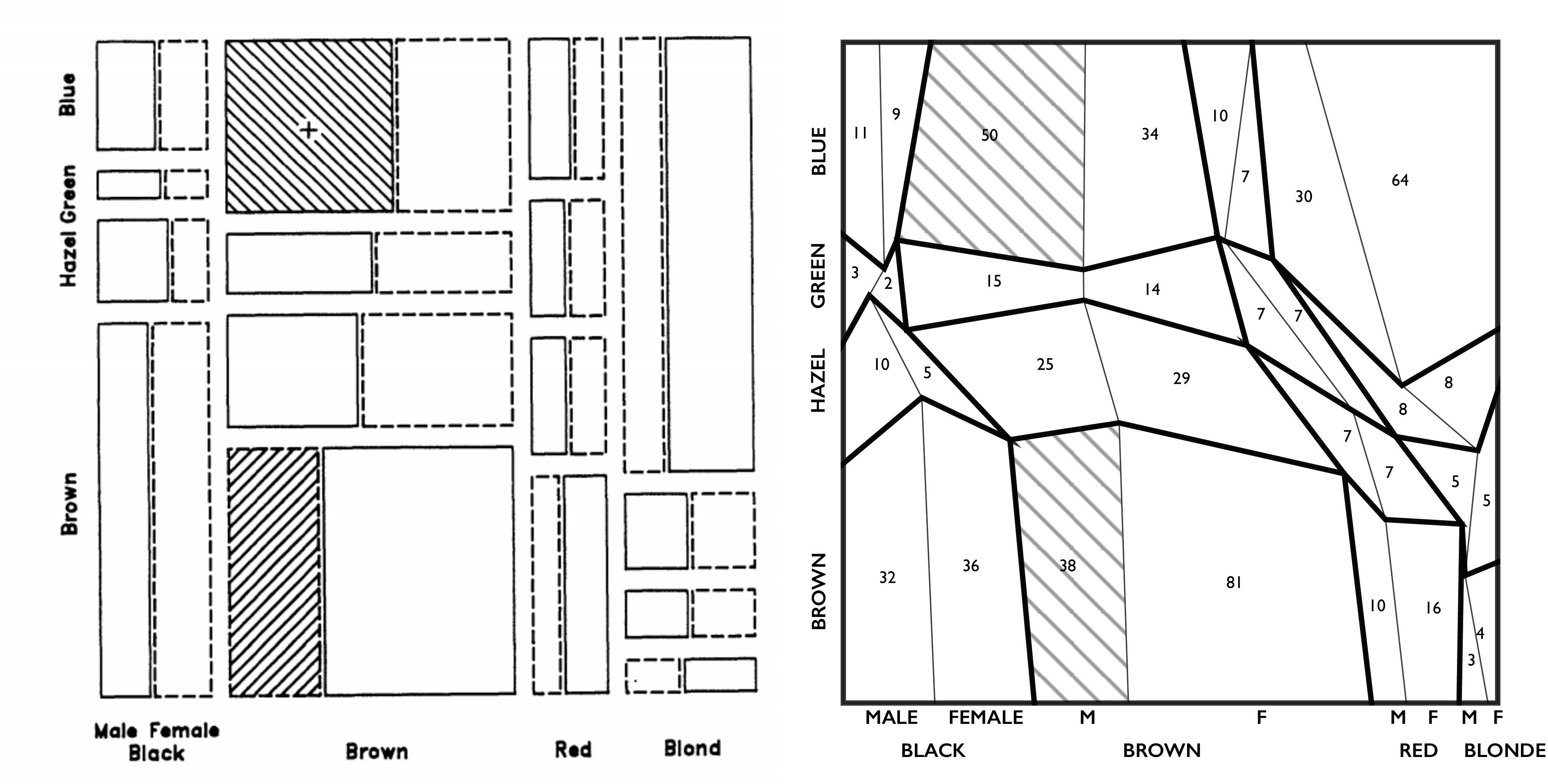}
    \caption{
        A comparison between a mosaic diagram created by Friendly \cite{friendly1994mosaic}, and the same data constructed as a \taco{}.
        We add bold outlines to the \taco{} to mirror the spaces that Friendly denotes with whitespace, and numbers (though they are not featured in the original image) for context.
        While the \taco{} rigidly maintains adjacency of the input data it does so at the expense of legibility, as comparisons of area in rectangles are easier to make than comparisons between \emph{``blobs''} (as the quadrilaterals of the \taco{} might be interpreted) \cite{cleveland1986experiment}.
        This exchange may be appropriate when the axes are ordinal, which is not the case here.
    }
    \label{fig:mosaic-compare}
    \spaceit
\end{figure}

In this supplement we provide a collection of asides and graphic studies demonstrating various \taco{} properties that did not fit in the main paper.
Among these are \figref{fig:multiplication} which provides a supplementary visual explanation of the layout \halluc{}, and \figref{fig:mosaic-compare} which compares presentations of a hierarchical nominal dataset.

\subsection{Non-axial Table Cartogram Arrangements}

\begin{figure*}[t] 
    \centering
    \includegraphics[width=\linewidth]{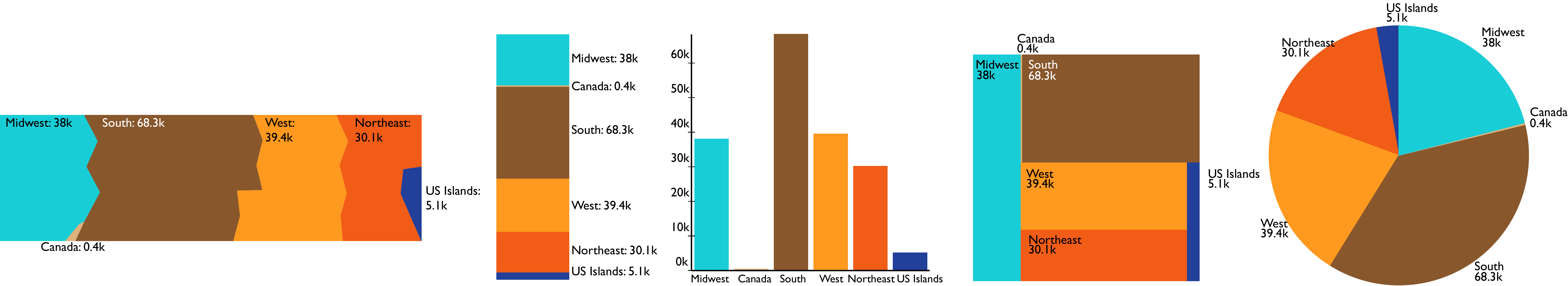}

    \caption{
        Bird-airplane collisions\cite{faa-strike} by region of flight origin rendered as a table cartogram by treating a waffle plot of original data as a table, as well as some common alternatives. (Left to right: \taco{}, stacked bar chart, bar chart, tree map, pie chart).
    }
    \label{fig:polygram}
    \spaceit

\end{figure*}

\begin{figure}[t]
    \centering
    \includegraphics[width=\columnwidth]{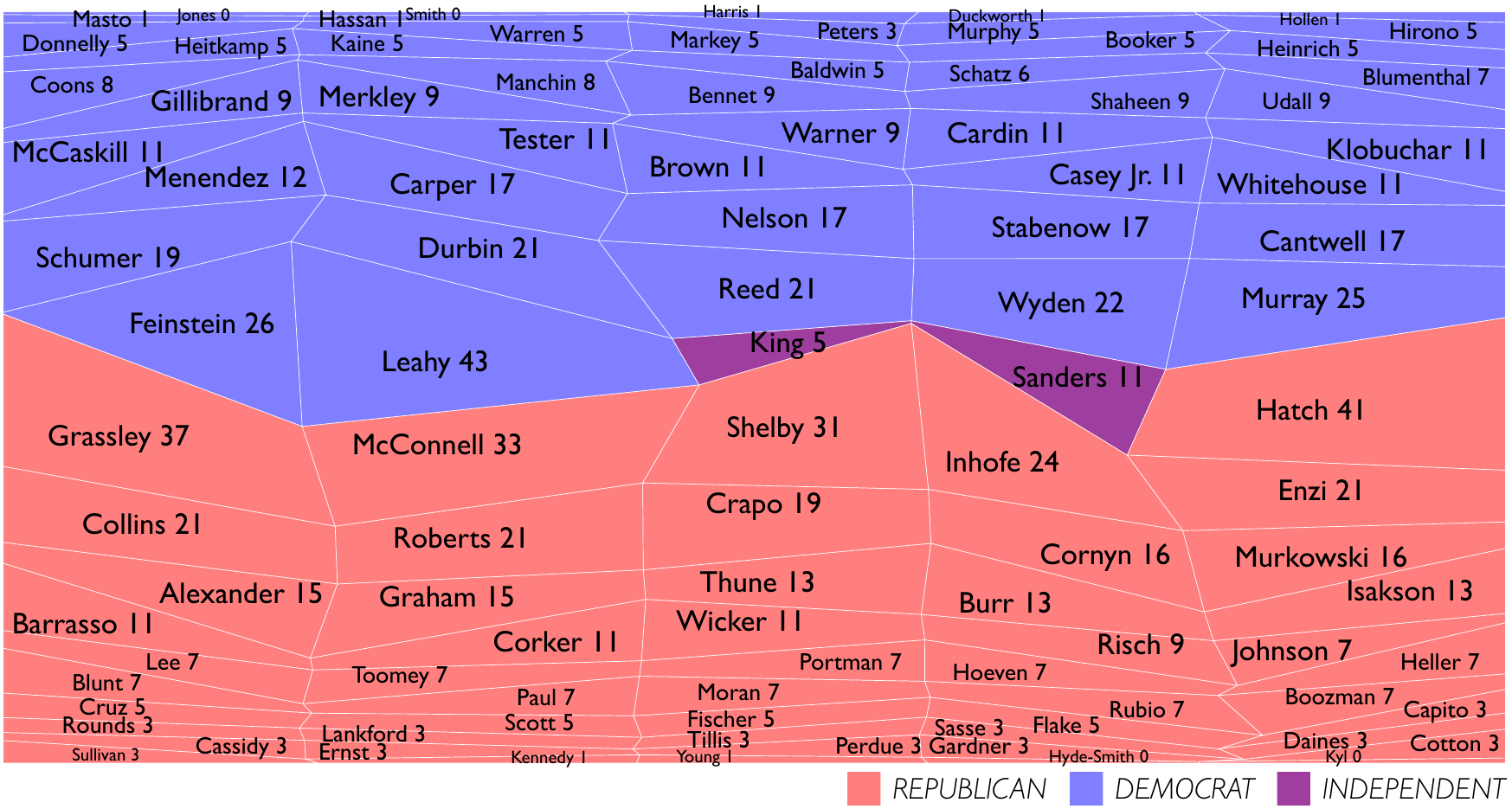}
    \caption{
        Political party and length of incumbency in US senate, fall 2018 \cite{senateWebsite}. Cells are organized into a waffle plot layout by party and by incumbency length and then mapped into a \taco{}.
    }
    \label{fig:senate}
    \spaceit
\end{figure}

Tabular renderings of data are often organized by the meaning of the axes, that is by rows and columns---however this need not always be the case. Such non-axial layouts can be rendered as table cartograms, as in \figref{fig:polygram} and \figref{fig:senate}. These possess a \halluc{}, as any arrangement of the data in a 2D layout will be non-reflective of the data and thus can be manipulated to make certain data classes appear larger or smaller than their actual data may warrant.
These structures abandon one of the key affordances of the table cartogram: the rigid maintenance of adjacency, which makes it substantially less clear why one might select this design over an equivalent one (such as  any of those shown in \figref{fig:polygram}).
We suggest that such structures might only be appropriate in cases where novelty is an essential design component.


\subsection{Area Embedding in Familiar Forms}

\begin{figure}[t] 
    \centering
    \includegraphics[width=\linewidth]{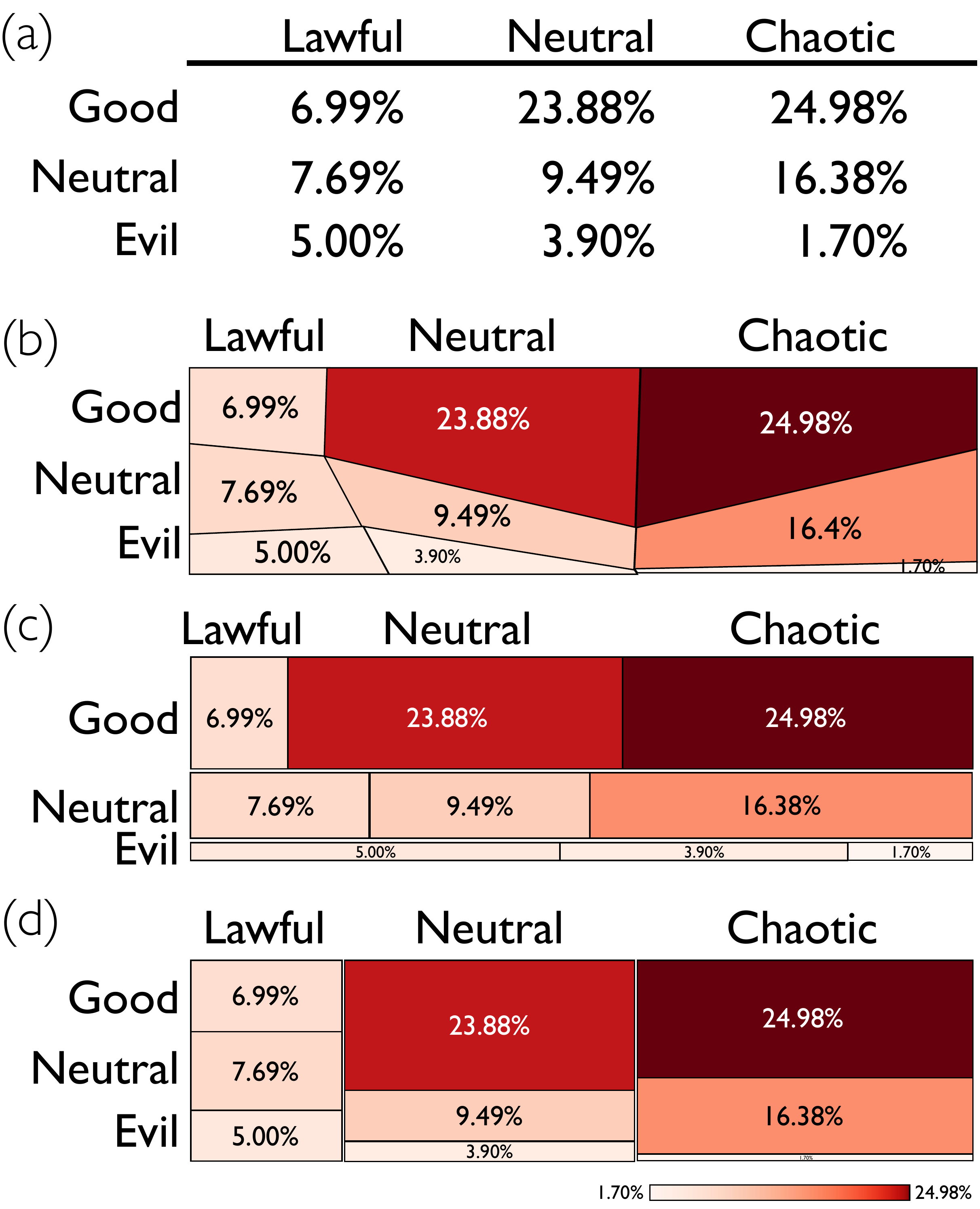}
    \caption{
        The popularity of the character alignments available to players in Dungeons and Dragons found in an informal online survey \cite{reddit:dnd-alignment}
        , across a standard table (a), a \taco{} (b), and two mosaics (c, d).
        While the table surfaces maxima via the number of digits, the other forms afford the same task (for both maxima and minima) via area.
        The \taco{} gives access to both marginal values, while the tree maps afford an accurate reading of a single marginal value (as comparing rectangular areas is more accurate than comparing blobs \cite{cleveland1986experiment}).
    }
    \label{fig:dnd}
    \vspace{-0.2in}
\end{figure}

Sometimes tabular data describes entities that already canonically inhabit a table, such as slices of the periodic table of elements (as exemplified by Evans \etal \cite{evans2013table}).
The reader's familiarity with that tabular layout simplifies understanding the table cartogram, analogous to how geographic cartograms leverage the reader's prior knowledge of geography.
\figref{fig:dnd} shows a table cartogram based on a Dungeons and Dragons \emph{moral alignment chart}, a standard and frequently satirized $3\times3$ table.
Readers familiar with that form can quickly read that players tend to prefer to be good and chaotic, since the visualization respects the standardized placement of those properties.
While a similar effect could be achieved using a slice-dice tree map or mosaic plot, it may have the effect of breaking from the canonical form as in \figref{fig:dnd}c.
This also aligns with our suggestion that calendar displays pair effectively with \taco{}s, as their form is well understood and commonly used.

\subsection{On Data with Zeros}

Rather than merely arguing data with zeros this should be avoided entirely, we now briefly consider a design strategy to address tabular data with values in $\mathbb{R}^{\geq0}$.
Despite recommendations to the contrary, a designer may feel that it is necessary to construct a table cartogram including such data.
In these cases a reasonable approach is to treat the data values as interval (as discussed in \secref{sec:data-value}, can be used if the task is appropriate, such as \task{Find Extremum} or \task{Retrieve Value}), as this allows for arbitrary shifts and rescaling to numerical data. The zero value can then be rescaled to a visually appropriate value, which will necessarily be design dependent.

An example of this strategy is Pierebean's analysis of frequency of Cantonese characters by pronunciation \cite{pierebean_oc_2020}. This data is organized into a table by international phonetic alphabet initials (rows) and finals (columns), with each cell made into a scalar by counting the number of characters that match that combination. After the table cartogram layout is computed, a word cloud of the characters in each cell is placed into each cell. There are a number of combinations that do not occur in the Cantonese language, yielding zeros. To address this the designer treated these counts as interval and offset them to give the desired form to the design (in particular setting the zero value to be 25).

While this selection is perhaps algebraically unsound---as ratio data should be treated as ratio data while interval data should be treated as interval data---it does serve a different goal of being visually interesting and giving granular access to the data.
Alternatives explored elsewhere in this supplement are unable to capture this effect. Mosaics would destroy the legibility of the axes.
A heatmap would not leave room for the literal representation of the characters (which appears to be an important part of the graphic's appeal).
Something more exotic, such as a gridded beeswarm chart would present the data as ratio and would still expose the literal encoding\cite{brath2020literal} but at expense of the vertical and horizontal space (though the Pierebean's graphic is already quite large).
Each of these alternatives should be considered before using a \taco{} to represent ratio data with zeros.
If these are insufficient then we recommend that the designer \emph{``interval-ize''} their data, use annotations and labels to prompt ordinal comparisons, and to annotate zeros through any available channel (such as texture or color).

The tension between this potentially problematic usage and design desire is exactly akin to the problem of specifying normalization conditions in dual-y-axis charts (which is a \halluc{}). We include this recommendation here because practitioners break formal rules (again just as in dual-y-axes), and we would prefer to give comprehensive guidance rather than piecemeal.

\begin{figure*}[ht]
    \centering
    \includegraphics[width=2\columnwidth]{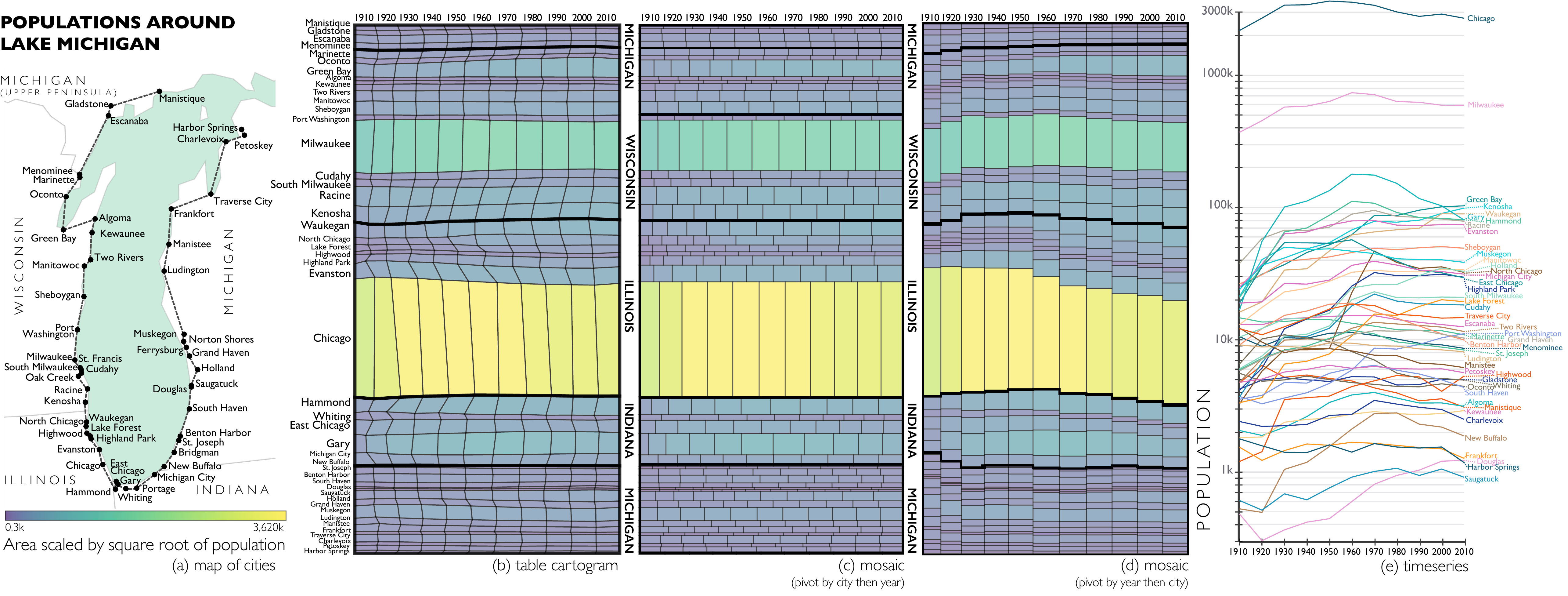}
    \caption{
        Several displays of changes in populations of cities around Lake Michigan during the 20th century. (b)-(d) utilize an ordering created by a counter-clockwise lake (a) traversal starting at the break between Michigan's upper and lower peninsulas.
    }
    \label{fig:teaser}
\end{figure*}

\begin{figure*}[t]
    \centering
    \includegraphics[width=2\columnwidth]{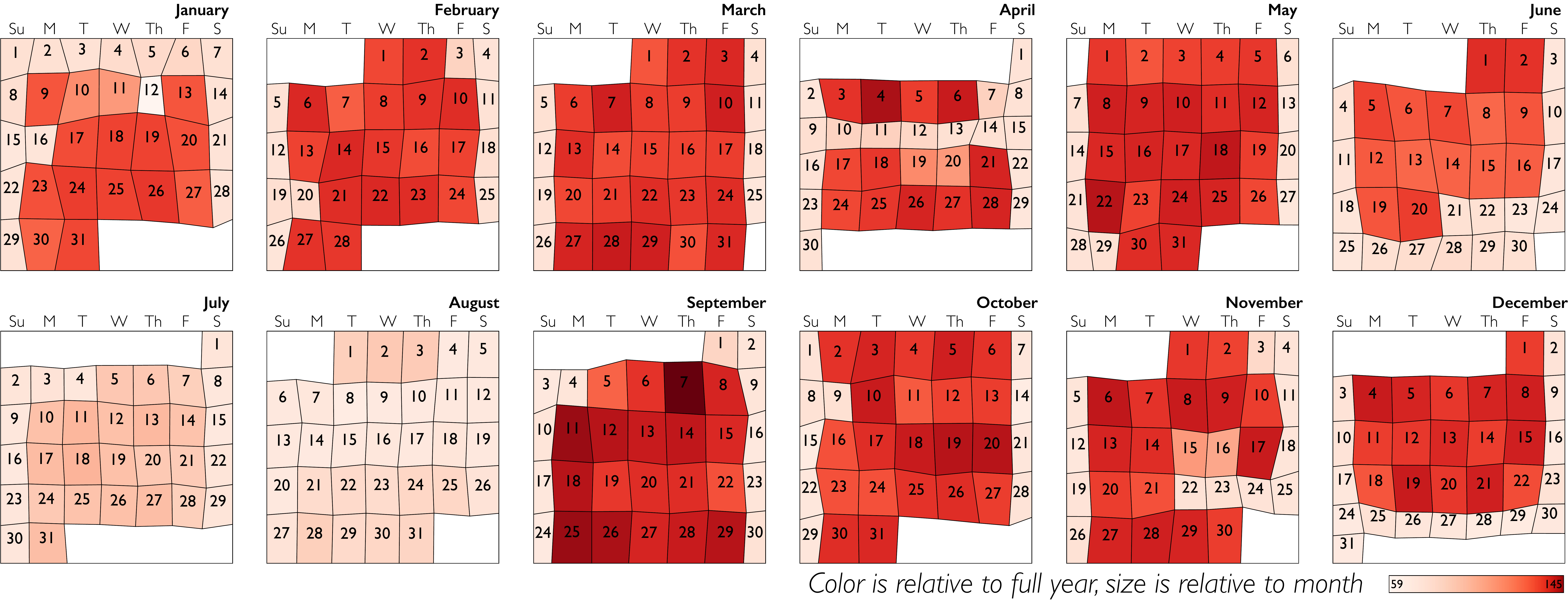}
    \caption{
        This visualization reveals how speed camera violations in Children's Safety Zones in Chicago\protect\cite{chicago-red-light} are correlated with the Chicago public schools calendar (e.g., the April 10-14 spring break accompanies an unusual dip in violations\cite{cps-schedule}).
    }
    \label{fig:year-calendar}
    \vspace{-0.2in}
\end{figure*}

\subsection{On Rendering Table Cartograms}

A possible factor for the current rarity of \taco{}s is a lack of easily accessible implementations.
At the time of this writing there are two known publicly available implementations, both of which utilize optimization-based schemes for computation \cite{Hasan2021, mcnutt2020Minimally}.
There are several more known versions, which are not available publicly, including Inoue and Li's \cite{inoue2020optimization} optimization-based approach, and Evans \etals \cite{evans2013table} geometry-based and optimization-based approaches. The latter of which they reference as producing their figures, although details are not provided.
The figures in this paper are created using our typescript implementation\cite{mcnutt2020Minimally}, as its  permissive interface design suited our need to explore the possible output space generated by the various parameter configurations.
In future work it would be interesting to compare the characteristics of the \taco{}s created by these implementations.

\subsection{Additional Design Studies}

\parahead{A larger calendar} As we saw in \secref{sec:case-study} year and month calendars can be usefully combined to create more complex compositions. In \figref{fig:year-calendar} we show a slightly simpler one,
that  reveals how these traffic violations are highly correlated with the Chicago public schools calendar (e.g., the April 10-14 spring break accompanies an unusual dip in violations\cite{cps-schedule}). \figref{fig:cal-diff} is drawn from this figure.
We already use calendars to structure our understanding of events and periods in the year, so just as other familiar forms may aid our understanding of graphics possessing an area-embedded structure, so too can these forms promote understanding of data presented in this manner.

\parahead{Spatio-temporal data}
We argued  that table cartograms are best suited to datasets with ordinal axes, which manifested as calendars in several examples. Yet this is by no-means the entirety of the possibilities for this  category of data.
In \figref{fig:teaser} we consider one such dataset: the populations of the cities around Lake Michigan over the 20th century\cite{census-pops}.
In (b-d) the ordering of the cities around the lake are linearized by traversing from counter-clockwise from the natural break at the top of Michigan (a), this is necessary to construct a relevant 1D ordering that might be used by one of the axes of a table cartogram.
(b-e) all facilitate basic summaries, like that Chicago always has been the most populous city on the lake, but (c-e) lose temporal or ordinal detail.
In (c) population changes over time are lost, in (d) the notion of a city is illegible due to the number of  cities being pivoted across, and in (e) the ordering of the cities around the lake is lost, which precludes making observations about the geographic distribution of populations.
The table cartogram (b) provides these properties by maintaining order and adjacency of the cells in the table.
However it does so at the price of precise comparison between years or cities (in d and c respectively). This might not be a significant deficit if the task prefers individual comparisons instead of marginal ones. These graphics might be interestingly combined through Ritchie \etals  \cite{ritchie2019lie} quasi-modes in which alternate views are presented on-demand.
This graphic features a minor \halluc{} in that the break-point that generates the linearization of the cities is arbitrarily chosen, as other values would lead to differing layouts.

\end{document}